\documentclass[revtex4,twocolappendix]{emulateapj}

\usepackage{amssymb}
\usepackage{amsmath}
\usepackage[]{graphicx}
\usepackage{enumerate}
\usepackage{subeqnarray}
\usepackage{cases}
\usepackage{mathrsfs,amssymb}
\usepackage[usenames]{color}

\tightenlines

\begin{document}

\title{Rigorous theory for secondary cosmic-ray ionization}

\author{Alexei V. Ivlev$^1$, Kedron Silsbee$^1$, Marco Padovani$^2$, Daniele Galli$^2$}
\email[e-mail:~]{ivlev@mpe.mpg.de} \affiliation{$^1$Max-Planck-Institut f\"ur Extraterrestrische Physik, 85748 Garching,
Germany } \affiliation{$^2$INAF--Osservatorio Astrofisico di Arcetri, Largo E. Fermi 5, 50125 Firenze, Italy}

\begin{abstract}
The energy spectrum of electrons produced in molecular gas by interstellar cosmic rays (CRs) is rigorously calculated as a
function of gas column density $N$ traversed by the CRs. This allows us to accurately compute the local value of the
secondary ionization rate of molecular hydrogen, $\zeta_{\rm sec}(N)$, as a function of the local primary ionization rate,
$\zeta_p(N)$. The ratio $\zeta_{\rm sec}/\zeta_p$ increases monotonically with $N$, and can considerably exceed the value of
$\approx0.67$ commonly adopted in the literature. For sufficiently soft interstellar spectra, the dependence $\zeta_{\rm
sec}/\zeta_p$ versus $N$ is practically insensitive to their particular shape and thus is a general characteristic of the
secondary CR ionization in dense gas.
\end{abstract}


\maketitle

\section{Introduction}

The ionization of dense gas by cosmic rays (CRs) is a problem of fundamental importance in astrophysics and astrochemistry.
Being the dominant source of ionization \citep{McKee1989, Caselli1998, Neufeld2017} and UV emission \citep{Prasad1983} in
dark regions, low-energy CRs govern the evolution of molecular clouds and the formation of stars \citep{Caselli2012,
Padovani2020}. The processes induced by CRs affect both the chemistry \citep{Keto2008, Keto2014} and thermodynamics
\citep{Galli2002, Glassgold2012, Ivlev2019} of the clouds. Furthermore, the level of ionization controls the degree to which
the gas is coupled to the magnetic field \citep{Shu1987}, which has profound implications for properties of disks around
young stars \citep{Zhao2016, Zhao2018}.

CRs interacting with the gas generate electron-ion pairs, with electrons having sufficient energy to produce further
ionization. These processes of primary and secondary ionization are characterized by the respective ionization rates (the
number of ionizations per unit time and per gas particle), $\zeta_p$ and $\zeta_{\rm sec}$. While $\zeta_p$ can be
straightforwardly derived for a given CR spectrum, computing $\zeta_{\rm sec}$ is a much more subtle task.
\citet{Dalgarno1958} first calculated the ratio $\zeta_{\rm sec}/\zeta_p$ for monoenergetic protons in atomic hydrogen,
finding a value of $\approx0.67$ for proton energies above few MeV. This value was later adopted by \citet{Spitzer1968} and
other authors as a constant multiplicative factor for an interstellar proton spectrum. For monoenergetic protons interacting
with molecular hydrogen, \citet{Glassgold1973} found $\zeta_{\rm sec}/\zeta_p$ increasing from 0.23 to 0.54 for energies
between 1~MeV and 10~MeV, while \citet{Cravens1978} reported ratios from 0.44 to 0.74 for energies between 1~MeV and
100~MeV.

In the present paper we rigorously compute the energy spectrum of electrons that are produced by interstellar CRs
penetrating dense astrophysical objects (such as molecular clouds or circumstellar disks), and derive the resulting rate of
secondary ionization as a function of the gas column density. We point out that knowing the exact spectrum of secondary
electrons makes it possible to accurately evaluate characteristics of other important processes driven by CRs, such as the
local rates of gas heating and H$_2$ dissociation, as well as the local magnitude of UV field due to H$_2$ fluorescence.

Unlike the approach by \citet[][]{Dalgarno1958} \citep[and similar approaches by][]{Knipp1953,Erskine1954}, aimed to
calculate the average number of ion pairs, we obtain a balance equation describing the steady-state electron spectrum, which
is similar to the degradation equation by \citet[][]{Spencer1954}. To the best of our knowledge, this is the first attempt
to accurately compute the secondary electron spectrum produced by CRs. Previous attempts \citep[e.g.,
by][]{Spencer1954,Xu1991} were focused on calculating the spectral degradation of monoenergetic electrons. Furthermore, most
astrophysical applications addressed the interaction of CRs with atomic or partially ionized low-density and low-column
density gas, while applications to dense gas neglected any dependence of $\zeta_{\rm sec}/\zeta_p$ on the column density. We
show that this ratio is not a constant, but increases with the column traversed by CRs, and that its magnitude can
considerably exceed the commonly adopted value of $\approx0.67$. It is worth noting that our approach can be easily
generalized to compute secondary X-ray ionization.

\section{Definitions and assumptions}
\label{def}

The energy distribution of CR species is characterized by their spectrum $j(E)$, which has the dimensions of a differential
flux per unit energy and solid angle (cm$^{-2}$~s$^{-1}$~eV$^{-1}$~sr$^{-1}$) and depends on the kinetic energy $E$. We are
interested in calculating the ionization rate of molecular hydrogen. The ionization is assumed to be due to interstellar CR
protons, the contribution of interstellar electrons is neglected (see discussion in Section~\ref{sec_balance}). Protons with
the local (attenuated) spectrum $j_{p}(E,N)$ produce primary ionization of H$_2$, occurring at the gas column density $N$ at
a rate of $\zeta_{p}(N)$. This generates {\it secondary electrons} with the local spectrum $j_{\rm sec}(E,N)$, leading to
secondary ionization of H$_2$ at a rate of $\zeta_{\rm sec}(N)$. Adding heavier CR nuclei with the interstellar spectrum
proportional to that of protons introduces a negligible contribution to the dependence $\zeta_{\rm sec}/\zeta_p$ versus $N$
(see Section~\ref{sec}). For the sake of clarity, $j_{\rm sec}$ is calculated neglecting ionization of helium and heavier
gas species, but our approach is applicable in general to arbitrary gas composition. The gas is assumed to be neutral,
because the effect of Coulomb collisions is vanishingly small for the ionization fractions expected in dense clouds (see
Section~\ref{Coulomb}).

We would like to stress that the definition of ``primary electrons'' adopted in literature often refers to the {\it first}
generation of electrons produced by CRs. In fact, the self-consistent treatment \citep{Spencer1954} does not make any
distinction between electron's generations, and therefore {\it all} produced electrons should be treated as secondary.

\subsection{Differential ionization cross sections}

The primary and secondary ionization of gas species is generally characterized by the respective differential cross
sections, $\partial \sigma_p/\partial \varepsilon$ and $\partial \sigma_e/\partial \varepsilon$, which are functions of $E$
and $\varepsilon$. The cross sections determine the probability that a proton ($p$) or electron ($e$) of energy $E$ produces
an ejected electron of energy $\varepsilon$.

For proton impact ionization we adopt the following approximate expression \citep[][]{Rudd1987,Rudd1988, Rudd1992}:
\begin{equation}\label{x-section_p}
\frac{\partial \sigma_p}{\partial \varepsilon}(E,\varepsilon)\approx\frac{f_p(E)}{(1+\tilde\varepsilon)^3}
\left[\tilde\varepsilon+\eta_p\ln\left(\frac{m_e}{m_p}\tilde E\right)\right],
\end{equation}
where $\tilde E=E/I$ and $\tilde \varepsilon=\varepsilon/I$ denote the energy normalization by the ionization potential $I$,
and $f_p(E)\propto E^{-1}$. The first term in the brackets represents the contribution of binary proton-electron collisions,
while the logarithmic term with the prefactor $\eta_p$ characterizes the dipole contribution from the Bethe theory
\citep[][]{Bethe1930, Landau1981_quantum}, arising due to dominant small-momentum transfer in inelastic collisions with a
molecule.

It must be stressed that in Equation~(\ref{x-section_p}) we use the expression valid for $\tilde E\gg m_p/m_e$, i.e., we
assume that the proton energy is much larger than $3\times10^4$~eV: as shown in Section~\ref{CR_spectrum}, the primary
ionization at column densities over $\sim10^{20}$~cm$^{-2}$ is determined by protons with energies much higher than this
value. Hence, the accuracy of Equation~(\ref{x-section_p}) is completely sufficient for the purposes of our studies.

The value of $\eta_p$ varies a little from one literature source to another. For the ionization of molecular hydrogen,
Equation~11 in \citet[][]{Rudd1987} gives 0.791/0.917=0.863; Equation~10 with Table~I in \citet[][]{Rudd1988} suggests
0.80/1.06=0.755; and Equations~43--48 with Table~V in \citet[][]{Rudd1992} gives 0.96/1.04=0.923. In this paper, we adopt
the latter value.

The differential cross section for the electron impact ionization takes into account exchange effects. In this case, we
generally write \citep[][]{Kim1994,Kim2000}
\begin{equation}\label{x-section_e}
\frac{\partial \sigma_e}{\partial \varepsilon}(E,\varepsilon)=f_e(E)\left[\varphi_{\rm M}(\varepsilon,\varepsilon')+
\eta_e\ln \tilde E\:\varphi_{\rm dip}(\varepsilon,\varepsilon')\right],
\end{equation}
where $\varepsilon$ and $\varepsilon'=E-\varepsilon-I$ are energies of two electrons produced by impact of an electron with
energy $E$. The function $f_e(E)$, given by the first factor of Equation~3 in \citet{Kim2000} multiplied by $2/(1+\eta_e)$,
varies as $f_e\propto E^{-1}$ for $E\gg I$. The first term in the brackets describes binary electron-electron collisions
according to the modified Mott's formula \citep[][]{Mott1930,Landau1981_quantum},
\begin{equation}\label{Mott}
\varphi_{\rm M}(\varepsilon,\varepsilon')=\frac1{(1+\tilde\varepsilon)^2}+\frac1{(1+\tilde\varepsilon')^2}-
\frac1{(1+\tilde\varepsilon)(1+\tilde\varepsilon')}\;.
\end{equation}
The function $\varphi_{\rm dip}(\varepsilon,\varepsilon')$ is determined by the differential dipole oscillator strength of a
molecule. For this paper, we use a symmetrized expression suggested by \citet[][]{Kim2000}
\begin{equation}\label{dip}
\varphi_{\rm dip}(\varepsilon,\varepsilon')=\frac1{(1+\tilde\varepsilon)^3}+\frac1{(1+\tilde\varepsilon')^3}\;.
\end{equation}
The prefactor $\eta_e=Q/(2-Q)$ is expressed via a dipole constant $Q$, a functional of the oscillator strength
\citep[see][]{Kim1994}. The simple form of Equation~(\ref{dip}) is suggested to use in cases where no reliable data on the
oscillator strength are available. Generally, $\varphi_{\rm dip}(\varepsilon,\varepsilon')$ is approximated by a
(symmetrized) polynomial of $(1+\tilde\varepsilon)^{-n}$ with $n\geq3$ \citep[][]{Kim1994,Kim2000}, which can be
straightforwardly included in our theory.

The value of $\eta_e$ appears to be less constrained than $\eta_p$. \citet[][]{Kim1994} and \citet[][]{Kim2000} suggest to
set $Q=1$ ($\eta_e=1$) when no data are available for a given gas species; at the same time, for hydrogen atoms they give
$Q=0.5668$ ($\eta_e\approx0.4$). On the other hand, our calculations in Section~\ref{sec} show that the ratio $\zeta_{\rm
sec}/\zeta_{p}$ is insensitive to $\eta_e$, and therefore we set $\eta_e=1$.

The use of relativistic expressions for the differential cross sections does not affect the principal results reported in
the paper. In particular, calculations with a relativistic formula for the electron impact ionization \citep{Kim2000} leave
almost unchanged the value of $\zeta_{\rm sec}/\zeta_{p}$ (see Section~\ref{sec}), leading to its slight increase by less
than 2\% at the largest analyzed column densities. Thus, for the sake of convenience we can employ non-relativistic
expressions (\ref{x-section_p}) and (\ref{x-section_e}), even though the high-energy tail of secondary electrons may become
relativistic for large columns.

\subsection{Ionization cross sections}

They are obtained by integrating the respective differential cross sections over a range of possible ejected energies. For
primary ionization, Equation~(\ref{x-section_p}) is integrated from 0 to
\begin{equation}\label{eps_max_p}
\varepsilon_{{\rm max},p}=4\frac{m_e}{m_p}E-I,
\end{equation}
the maximum energy that can be transferred by a proton of energy $E$ to the ejected electron. Since
Equation~(\ref{x-section_p}) is applicable for proton energies such that $\varepsilon_{{\rm max},p}\gg I$, we extend the
integration to infinity (thus omitting terms going beyond the assumed applicability). This gives the following approximate
expression:
\begin{equation}\label{sigma_ion_p}
\sigma_{{\rm ion},p}(E)\approx\frac12f_p(E)I\left[1+\eta_p\ln\left(\frac{m_e}{m_p}\tilde E\right)\right],
\end{equation}
valid for $\tilde E\gg m_p/m_e$. For the electron impact ionization, we integrate Equation~(\ref{x-section_e}) up to
\begin{equation}\label{eps_max_e}
\varepsilon_{{\rm max},e}=\frac12(E-I),
\end{equation}
the maximum value of $\varepsilon$ for indistinguishable electrons. This yields the following general formula:
\begin{eqnarray}
\sigma_{{\rm ion},e}(E)=f_e(E)I\bigg[1-\frac1{\tilde E}-\frac{\ln \tilde E}{1+\tilde E}\hspace{2.5cm}\label{sigma_ion_e}\\
+\frac{\eta_e}2\left(1-\frac1{\tilde E^2}\right)\ln\tilde E\bigg],\nonumber
\end{eqnarray}
valid for any $E\geq I$.

\section{Local spectrum of CR protons}
\label{CR_spectrum}

Let us start with rigorous derivation of the steady-state kinetic equation for CR protons. Assuming their free-streaming
propagation \citep[see][and references therein]{Padovani2020}, the {\it local} spectrum $j_{p}(E,N,\mu)$ of protons with
pitch-angle cosine $\mu$ at column density $N$ is determined by a balance of advection and energy losses:
\begin{equation}\label{protons}
\mu\:\frac{\partial j_{p}}{\partial N}+\mathcal{P}-\mathcal{D}=0.
\end{equation}
The rates $\mathcal{P}(E)$ and $\mathcal{D}(E)$ at which ionizing collisions of protons lead, respectively, to population
and depopulation of their energy state $E$ (we do not indicate dependence on $N$ and $\mu$ for brevity) have the form
introduced by \citet{Fano1953} and \citet{Spencer1954}:
\begin{eqnarray}
\mathcal{P}(E)=\int_0^{\varepsilon_{{\rm max},p}^*}\frac{\partial \sigma_p}{\partial \varepsilon}(E+\varepsilon+I,\varepsilon)
j_{p}(E+\varepsilon+I)\:d\varepsilon,\hspace{.4cm}\label{S_pr}\\[.3cm]
\mathcal{D}(E)=j_{p}(E)\int_0^{\varepsilon_{{\rm max},p}}\frac{\partial \sigma_p}{\partial \varepsilon}(E,\varepsilon)\:
d\varepsilon\hspace{2.9cm}\label{L_pr}\\
\equiv \sigma_{{\rm ion},p}(E)j_{p}(E),\nonumber
\end{eqnarray}
where $\varepsilon_{{\rm max},p}$ is given by Equation~(\ref{eps_max_p}), while $\varepsilon_{{\rm max},p}^*$ is obtained
from Equation~(\ref{eps_max_p}) by replacing $E$ with $E+\varepsilon_{{\rm max},p}^*+I$. We note that pitch angles of
protons remain practically unchanged after ionizing collisions, and therefore Equations~(\ref{S_pr}) and (\ref{L_pr})
involve only integration over $\varepsilon$. In Appendix~\ref{A1} we show that, due to the presence of small parameter
$4(m_e/m_p)$, the difference $\mathcal{P}- \mathcal{D}$ can be written in a differential form. With the accuracy
$O(m_e/m_p)$, this leads to the standard kinetic equation in the continuous slowing-down approximation
\citep[e.g.,][]{Fano1953,Padovani2018},
\begin{equation}\label{protons2}
\mu\:\frac{\partial j_{p}}{\partial N}+\frac{\partial}{\partial E}\left(L_pj_{p}\right)\approx0,
\end{equation}
where $L_p(E)$ is the ionization loss function of protons, given by Equation~(\ref{L_p2}).

In fact, the proton spectrum is attenuated due to ionization and other mechanisms of continuous losses (such as excitation),
and then different contributions simply sum up in Equation~(\ref{protons2}). For a gas composed of multiple species, the
loss function is a sum of the respective partial contributions.

Equation~(\ref{protons2}) can be generally solved by the method of characteristics. The solution is determined by the proton
stopping range,
\begin{equation}\label{range0}
R_p(E)=\int_0^E\frac{dE'}{L_p(E')}\;,
\end{equation}
and can be explicitly derived for a power-law form of the interstellar (isotropic) spectrum, $j_{p}^{\rm IS}(E)$, see
Appendix~\ref{A2}. An important parameter applied in the analysis below is the proton attenuation energy $E_{\rm att}(N)$,
which is the inverse function of the stopping range,\footnote{Stopping ranges of different CR species are plotted in
Figure~2 of \citet[][]{Padovani2018}.}
\begin{equation}\label{range}
R_p(E_{\rm att})=N.
\end{equation}
Using a power-law approximation for the loss function of protons, Equation~(\ref{L_p}), their attenuation energy for
$10^{20}$~cm$^{-2}\lesssim N\lesssim10^{25}$~cm$^{-2}$ is approximated to within 2\% by
\begin{equation}\label{E_att}
E_{\rm att}(N)\approx 2.2\:N_{21}^{0.55}~{\rm MeV},
\end{equation}
where $N_{21}$ is the gas column density in units of $10^{21}$~cm$^{-2}$. Equation~(\ref{E_att}) is obtained assuming the
ISM composition by \citet{Wilms2000} with hydrogen in the molecular form. Here and below, $N$ denotes the column density of
all gas species, related to the H$_2$ column density via $N\approx1.20N_{\rm H2}$.

In what follows, the local spectrum of CR protons is calculated from the continuous slowing-down approximation,
Equation~(\ref{protons2}). \citet{Padovani2018} showed that this approximation becomes increasingly inaccurate around the
column density of $N=10^{25}$~cm$^{-2}$ and above, due to the growing effect of nuclear collisions accompanied by pion
production; therefore, in the present paper the maximum column used for calculations is set to this value. For the
interstellar spectrum, we assume a model form suggested by \citet[][]{Padovani2018},
\begin{equation}\label{j_HL}
j_{p}^{\rm IS}(E)=C\frac{E^{-a}}{(E+E_0)^b}~{\rm cm}^{-2}~{\rm s}^{-1}~{\rm eV}^{-1}~{\rm sr}^{-1},
\end{equation}
with $C=2.4\times10^{15}$ and $E_0=650$~MeV. Two characteristic models are considered: a ``high'' (soft) spectrum
$\mathscr{H}$ with $a=0.8$ and a ``low'' (hard) spectrum $\mathscr{L}$ with $a=-0.1$, both having the same high-energy
asymptote with $a+b=2.7$. The spectrum $\mathscr{H}$ has been previously introduced to fit available data on H$_2$
ionization in diffuse clouds \citep{Padovani2018}, while the spectrum $\mathscr{L}$ represents the proton spectrum measured
down to $E=3$~MeV by the Voyager~1 spacecraft \citep{Cummings2016} and extrapolated to the lower energies with the constant
slope \citep{Padovani2018}.

\section{Balance equation for the electron spectrum}

\begin{figure*}%[!h]
\begin{center}
\includegraphics[width=0.497\textwidth]{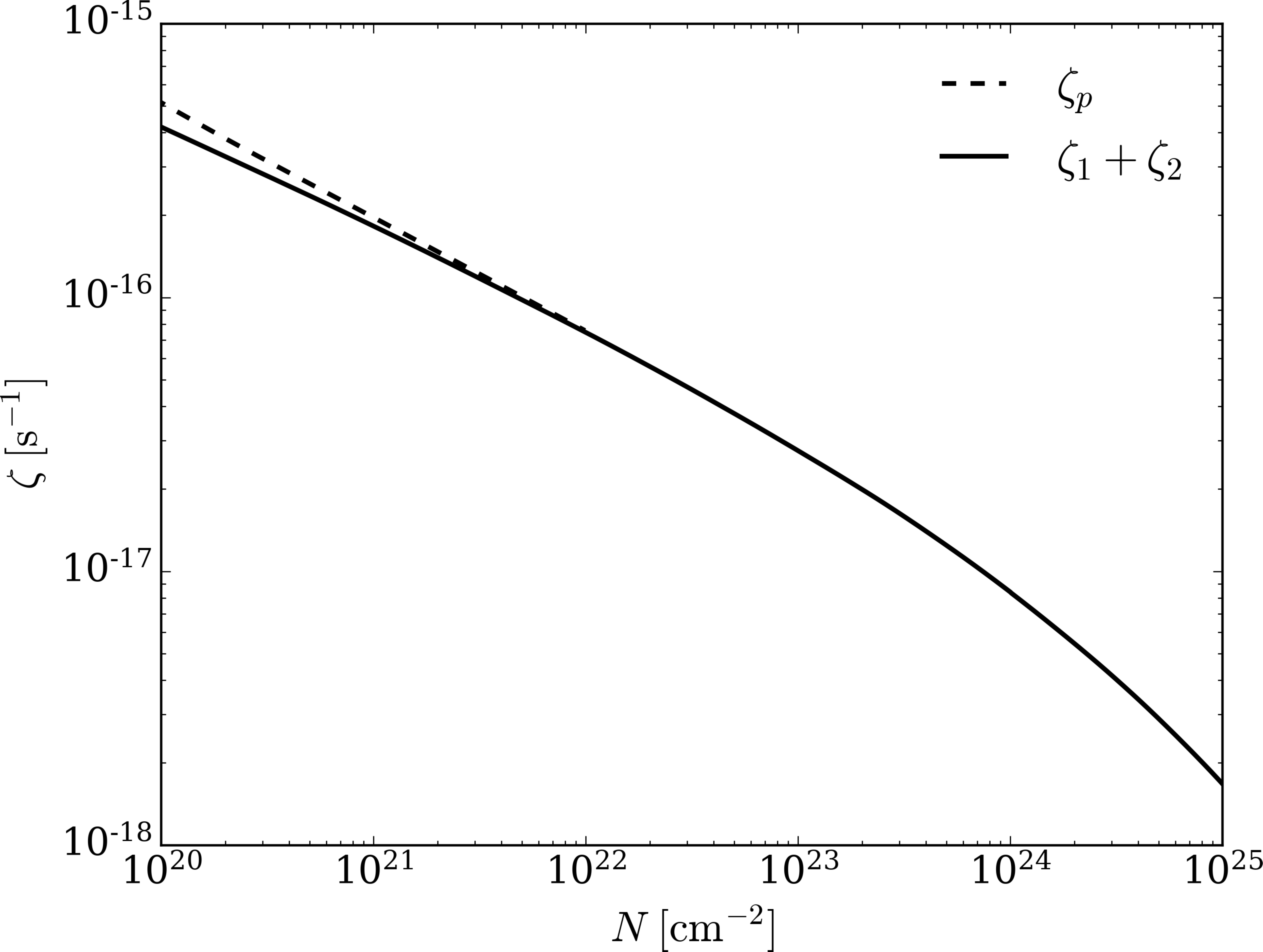}
\includegraphics[width=0.497\textwidth]{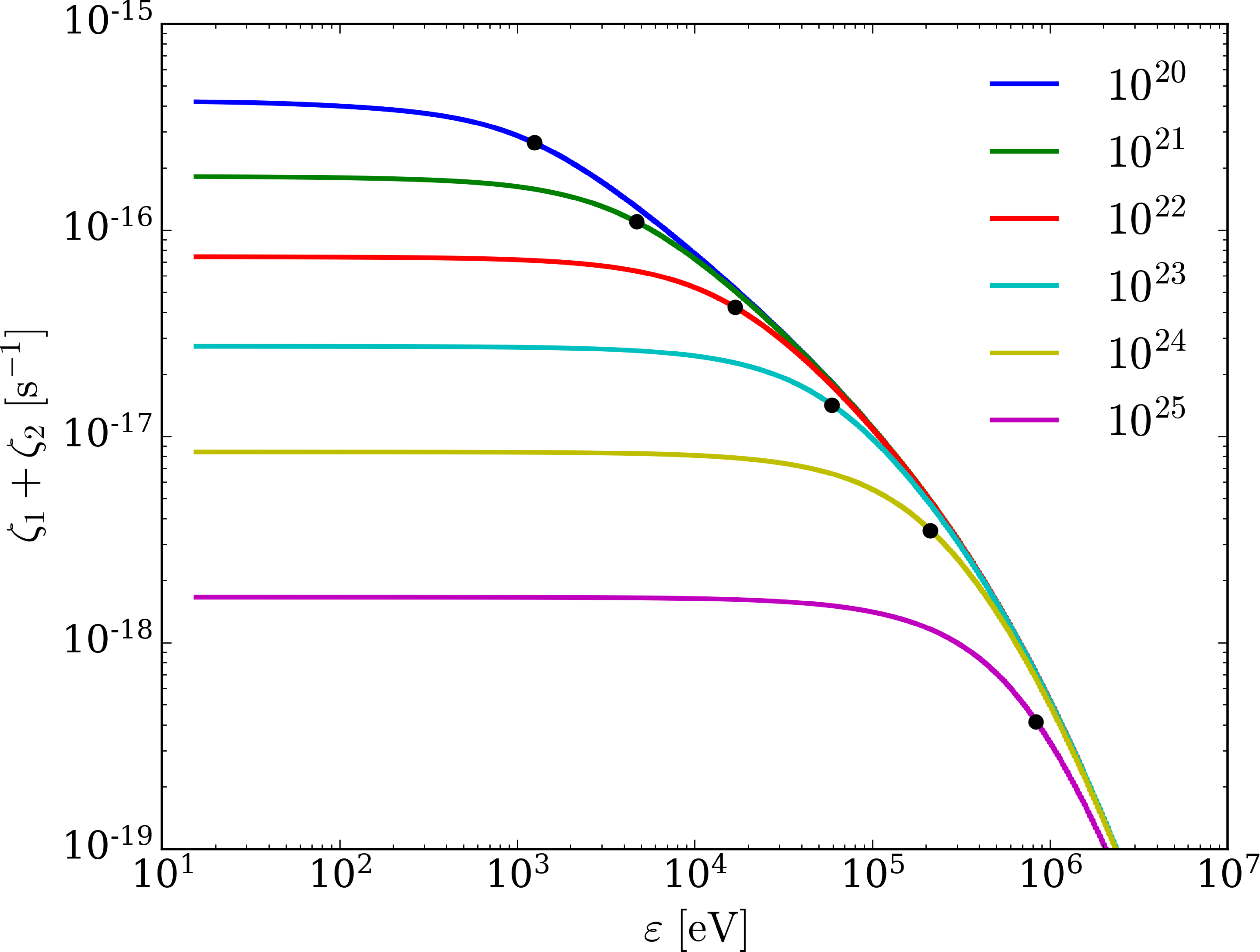}
\caption{Left panel: Rates of primary ionization of H$_2$ versus the gas column density $N$, computed using the continuous
slowing-down approximation (see Section~\ref{CR_spectrum}) for the interstellar proton spectrum $\mathscr{H}$. The dashed
line shows the dependence $\zeta_{p}(N)$ derived for the exact cross section of proton impact ionization \citep{Rudd1992},
the solid line represents the sum $\zeta_1+\zeta_2$ obtained from Equations~(\ref{zeta_1}) and (\ref{zeta_2}) for
$\varepsilon=I$. Right panel: $\zeta_1+\zeta_2$ versus $\varepsilon$ for different values of $N$ (see the legend, in units
of cm$^{-2}$). At smaller $\varepsilon$, each curve converges to the value shown by the solid line in the left panel. The
bullets indicate where $\varepsilon= \varepsilon_{\rm att}(N)$.} \label{fig_zeta12}
\end{center}
\end{figure*}

Unlike protons, the continuous slowing-down approximation is not applicable for electrons. Below we show that the difference
of the population and depopulation rates for secondary ionization (see Section~\ref{sec_secondary}) {\it cannot} be
presented in a differential form, as the energy exchange in such collisions is not small, and the electron
indistinguishability leaves a non-negligible integral term (see Appendix~\ref{A3}).

At the same time, transport of secondary electrons is negligible. Indeed, Equation~(\ref{j_CR}) in Appendix~\ref{A2}
suggests that the local proton spectrum (determining primary ionization) varies at a column scale of $\sim N$ for $E\lesssim
E_{\rm att}(N)$, and remains approximately constant for higher $E$. The fact that the ionization (and hence also excitation)
loss functions of electrons and protons, $L_e(E_e)$ and $L_p(E_p)$, respectively, are comparable for equal particle
velocities allows us to write the relation $L_e/L_p\sim(m_e/m_p)(E_p/E_e)$, valid with logarithmic accuracy for $E_e\gg I$
and $E_p\gg(m_p/m_e)I$. With the same accuracy, from Equation~(\ref{range0}) we derive $L_e/L_p\sim (E_e/E_p)(R_p/R_e)$, and
combining it with the preceding relation obtain $R_e/R_p\sim(m_p/m_e)(E_e/E_p)^2$. Since $E_e\lesssim4(m_e/m_p) E_p$,
substituting Equation~(\ref{range}) yields $R_e/N\sim 10m_e/m_p\sim0.01$ for the maximum stopping range of electrons
produced at a given column density. Therefore, we can safely assume that secondary electrons are attenuated locally.

Thus, the steady-state spectrum $j_{\rm sec}(\varepsilon)$ is governed by the {\it local} balance of primary ionization and
various loss mechanisms. Below we derive the balance equation for $j_{\rm sec}(\varepsilon)$, considering the secondary
ionization and excitation as the major loss processes.

\subsection{Primary ionization}
\label{primary}

Consider the production of secondary electrons upon the proton impact ionization of the gas. The source term due to the
primary ionization at given column density $N$, viz., the number of electrons produced at energy $\varepsilon$ (per unit
time per gas particle) by CR protons with the local spectrum $j_{p}(E,N)$, is
\begin{equation}\label{S_p}
\mathcal{P}_{p}(\varepsilon,N)=\int_{\frac14\frac{m_p}{m_e}(\varepsilon+I)}^{\infty}\frac{\partial \sigma_p}{\partial
\varepsilon}(E,\varepsilon)j_{p}(E,N)\:dE.
\end{equation}
To take into account the fact that the attenuation of interstellar CRs generally creates anisotropy with respect to the
magnetic field lines (see Section~\ref{CR_spectrum}), the CR spectrum in Equation~(\ref{S_p}) is averaged over the pitch
angles, i.e.,
\begin{equation}\label{j_av}
j_{p}(E,N)=\frac12\int_{-1}^1 j_{p}(E,N,\mu)\: d\mu.
\end{equation}
Assuming the integral over $E$ to be dominated by proton energies much larger than $\frac14(m_p/m_e)I\sim10^4$~eV, we can
substitute Equation~(\ref{x-section_p}) in Equation~(\ref{S_p}) and present the latter in the following form:
\begin{equation}\label{S_p1}
\mathcal{P}_{p}(\varepsilon,N)=\frac1{2\pi(1+\tilde\varepsilon)^3I}\left[\zeta_1(\varepsilon,N)+
\tilde\varepsilon\zeta_2(\varepsilon,N)\right],
\end{equation}
where the rates
\begin{eqnarray}
  \zeta_1(\varepsilon,N)\hspace{7cm}\label{zeta_1}\\
  =2\pi\eta_p I\int_{\frac14\frac{m_p}{m_e}(\varepsilon+I)}^{\infty}f_p(E)
  \ln\left(\frac{m_e}{m_p}\tilde E\right)j_{p}(E,N)\:dE,\nonumber\\[.3cm]
  \zeta_2(\varepsilon,N) =2\pi I\int_{\frac14\frac{m_p}{m_e}(\varepsilon+I)}^{\infty}f_p(E)j_{p}(E,N)\:dE,\hspace{1.1cm}
  \label{zeta_2}
\end{eqnarray}
are functionals of the local proton spectrum.\footnote{For the assumed values of $E$ we can omit $I$ in the lower
integration limit.} Thus, $\zeta_1$ and $\zeta_2$ determine the magnitude of the source term at low and high $\varepsilon$,
respectively.

For the further analysis, it is convenient to introduce the electron energy scale $\varepsilon_{\rm att}$, related to the
proton attenuation energy (\ref{E_att}) via
\begin{equation}\label{epsilon_att}
\varepsilon_{\rm att}(N)=4\frac{m_e}{m_p}E_{\rm att}(N)\approx 4.8\:N_{\rm 21}^{0.55}~{\rm keV}.
\end{equation}
Taking into account Equation~(\ref{sigma_ion_p}), the sum $\zeta_1+\zeta_2$ is the integral of the product $4\pi\sigma_{{\rm
ion},p}(E)j_{p}(E,N)$. Hence, for small $\varepsilon$ it tends to the actual rate of the local primary ionization,
$\zeta_{p}(N)$. Equation~(\ref{j_CR}) shows that $j_{p}(E,N)$ is peaked at $E\sim E_{\rm att}(N)$, and therefore the sum
remains independent of $\varepsilon$ and equal to $\zeta_{p}(N)$ for $\varepsilon\ll \varepsilon_{\rm att}(N)$. The latter
is demonstrated in the left panel of Figure~\ref{fig_zeta12}, obtained for the interstellar proton spectrum $\mathscr{H}$:
here, $\zeta_1+\zeta_2$ computed for $\varepsilon=I$ is plotted versus $N$ along with the dependence $\zeta_{p}(N)$ derived
from a precise expression for the ionization cross section \citep{Rudd1992}. The two curves nearly coincide for $N\gtrsim
10^{21}$~cm$^{-2}$, showing that the solid line is expected to accurately represent $\zeta_{p}(N)$ even for diffuse
envelopes of molecular clouds. As expected -- see discussion after Equation~(\ref{x-section_p}) -- a noticeable deviation is
only seen around $N\sim 10^{20}$~cm$^{-2}$, where $E_{\rm att}(N)\sim10(m_p/m_e)I$ and hence the adopted $\sigma_{{\rm
ion},p}(E)$ becomes slightly inaccurate.

For $\varepsilon\gtrsim \varepsilon_{\rm att}(N)$, both $\zeta_1$ and $\zeta_2$ become asymptotically independent of $N$.
Given $f_p(E)\propto E^{-1}$, they fall off with $\varepsilon$ approximately as $\zeta_1(\varepsilon)\propto j_{p}^{\rm
IS}(\frac14 \frac{m_p}{m_e}\varepsilon) \ln\tilde\varepsilon$ and $\zeta_2(\varepsilon)\propto j_{p}^{\rm IS}(\frac14
\frac{m_p}{m_e} \varepsilon)$, as determined by the form of the interstellar spectrum. The right panel of
Figure~\ref{fig_zeta12} illustrates this behavior for the interstellar spectrum $\mathscr{H}$. Here, $\zeta_1+\zeta_2$ is
plotted versus $\varepsilon$ for different values of $N$, showing how individual curves approach a common decreasing
asymptote at $\varepsilon\gtrsim \varepsilon_{\rm att}(N)$ and tend to the plateau $\approx\zeta_{p}(N)$ at lower
$\varepsilon$.

\subsection{Secondary ionization}
\label{sec_secondary}

The rate $\mathcal{P}_{\rm sec}(\varepsilon)$ at which secondary ionization collisions contribute to population of electrons
with energy $\varepsilon$ (at given $N$) can be easily calculated using Equations~(\ref{x-section_e})--(\ref{dip}). Setting
the energy of colliding electron to $E=\varepsilon+\varepsilon'+I$ and integrating the product $\partial \sigma_e/\partial
\varepsilon(E,\varepsilon) \:j_{\rm sec}(E)$ over $\varepsilon'$, we obtain
\begin{eqnarray}
4\pi I \mathcal{P}_{\rm sec}(\varepsilon)=\int_0^\infty \varphi_{\rm M}(\varepsilon,\varepsilon')F(\varepsilon+\varepsilon'+I)
\:d\varepsilon'\hspace{1.5cm}\label{S_s}\\
+\eta_e\int_0^\infty \varphi_{\rm dip}(\varepsilon,\varepsilon')\ln(\tilde\varepsilon+\tilde\varepsilon'+1)
F(\varepsilon+\varepsilon'+I)\:d\varepsilon'.\nonumber
\end{eqnarray}
Here, to simplify the presentation of the results in the following text, we added the factor $4\pi I$ in order to introduce
an auxiliary function $F$ for the secondary spectrum (of dimensions eV$^{-1}$~s$^{-1}$),
\begin{equation}\label{F_aux}
F(\varepsilon)\equiv 4\pi If_e(\varepsilon)j_{\rm sec}(\varepsilon).
\end{equation}
The rate of depopulation, $\mathcal{D}_{\rm sec}(\varepsilon)$, is simply equal to
\begin{equation}\label{L_s}
\mathcal{D}_{\rm sec}(\varepsilon)=\sigma_{{\rm ion},e}(\varepsilon)j_{\rm sec}(\varepsilon).
\end{equation}

\subsection{Excitation}

Consider electron collisions leading to excitation of state $k$ of a molecule, characterized by the excitation energy
$\Delta_k$. The difference of the corresponding population and depopulation rates is
\begin{eqnarray}
\mathcal{P}_{{\rm exc},k}(\varepsilon)-\mathcal{D}_{{\rm exc},k}(\varepsilon)\hspace{5cm}\label{S-L_exc}\\[.2cm]
=\sigma_{{\rm exc},k}(\varepsilon+\Delta_k)
j_{\rm sec}(\varepsilon+\Delta_k)-\sigma_{{\rm exc},k}(\varepsilon)j_{\rm sec}(\varepsilon),\nonumber
\end{eqnarray}
where $\sigma_{{\rm exc},k}(\varepsilon)$ is the excitation cross section of state $k$ \citep[see][and references
therein]{Dalgarno1999}.

\subsection{Balance equation}
\label{sec_balance}

By summing up different contributions to the population and depopulation rates of electrons with energy $\varepsilon$, we
obtain the following balance equation for the spectrum of secondary electrons:
\begin{widetext}
\begin{eqnarray}
\frac2{(1+\tilde\varepsilon)^3}\left[\zeta_1(\varepsilon,N)+\tilde\varepsilon\zeta_2(\varepsilon,N)\right]
\hspace{13.cm}\label{balance}\\
+\int_0^\infty \varphi_{\rm M}(\varepsilon,\varepsilon')F(\varepsilon+\varepsilon'+I)\:d\varepsilon'+\eta_e\int_0^\infty
\varphi_{\rm dip}(\varepsilon,\varepsilon')\ln(\tilde\varepsilon+\tilde\varepsilon'+1)F(\varepsilon+\varepsilon'+I)
\:d\varepsilon' +I\sum_k\Phi_{{\rm exc},k}(\varepsilon+\Delta_k)F(\varepsilon+\Delta_k)
\hspace{0.cm}\nonumber\\
=I\left[\Phi(\varepsilon)+\sum_k\Phi_{{\rm exc},k}(\varepsilon)\right]F(\varepsilon),
\nonumber
\end{eqnarray}
\end{widetext}
where dimensionless auxiliary functions for the ionization and excitation cross sections are
\begin{equation*}
\Phi(\varepsilon)=\frac{\sigma_{{\rm ion},e}(\varepsilon)}{f_e(\varepsilon)I}\quad {\rm and} \quad
\Phi_{{\rm exc},k}(\varepsilon)=\frac{\sigma_{{\rm exc},k}(\varepsilon)}{f_e(\varepsilon)I}\;,
\end{equation*}
respectively, and the dependence of $F$ on $N$ is not indicated for brevity.

Equation~(\ref{balance}) assumes collisions with the most abundant gas species, i.e., with hydrogen molecules. Collisions
with He and other gas species can be straightforwardly included by adding the corresponding terms (primary and secondary
ionization plus excitation) multiplied by the species abundance. In principle, a contribution of interstellar CR electrons
could also be included: this does not change the structure of Equation~(\ref{balance}), since interstellar and secondary
electrons are indistinguishable. On the other hand, it results in additional advection term (analogous to the first term in
Equation~(\ref{protons2}) for protons) and thus makes a solution of the balance equation much more complicated. However,
according to \citet{Padovani2018} the primary ionization is believed to be completely controlled by CR protons if their
spectrum is close to the model form $\mathscr{H}$ (for the spectrum $\mathscr{L}$ it is true for $N\gtrsim
10^{22}$~cm$^{-2}$), and therefore we neglect the effect of interstellar electrons in this paper.

Finally, we note that Equation~(\ref{balance}) can be easily generalized to compute the secondary electron spectrum produced
by X rays. In this case, $j_p$ and $\partial \sigma_p/\partial \varepsilon$ in the source term due to proton ionization,
Equation~(\ref{S_p}), are replaced by the corresponding X-ray spectrum and differential cross section, while the lower limit
of integration over the X-ray energy is $\varepsilon +I$. This only leads to a different functional form of the first term
in Equation~(\ref{balance}).

\subsection{Effect of Coulomb collisions}
\label{Coulomb}

The Coulomb collisions with free electrons could be included in Equation~(\ref{balance}), too, by adding the corresponding
rates multiplied by the gas ionization fraction $n_e/n_{\rm gas}$. The population rate is given by the first integral in
Equation~(\ref{S_s}) with $I=0$ and $\varphi_{\rm M}(\varepsilon,\varepsilon')$ described by classical Mott's formula for
free electrons \citep{Landau1981_quantum}, i.e., by Equation~(\ref{Mott}) without unity in the denominators; the
depopulation rate is proportional to the integral over this $\varphi_{\rm M}(\varepsilon,\varepsilon')$.

Obviously, the resulting integrals contain terms diverging as $\propto1/\varepsilon'$ at $\varepsilon'\to0$. This artificial
divergence is avoided in the balance equation by writing the difference of the population and depopulation rates as
$I^2\ln(\varepsilon/\varepsilon_{\rm min})F'(\varepsilon)$ plus non-diverging terms; here $F'(\varepsilon)$ denotes a
derivative over $\varepsilon$ and the factor $I^2$ comes from the common energy normalization. The minimum truncation energy
$\varepsilon_{\rm min}\sim(\delta p_{\rm min})^2/2m_e$ is determined by the minimum momentum $\delta p_{\rm min}\sim
e^2/(b_{\rm max}v)$ that can be transferred by a secondary electron (with the velocity $v=\sqrt{2\varepsilon/m_e}$\:) to the
surrounding free electrons (whose plasma frequency is $\omega_{{\rm p}e}=\sqrt{4\pi e^2n_e/m_e}$\:) at the maximum impact
parameter $b_{\rm max}\sim v/\omega_{{\rm p}e}$. The resulting logarithmic factor $\ln(\varepsilon/\varepsilon_{\rm
min})\sim3\ln(\varepsilon/e^2n_e^{1/3})$ is estimated to be $\lesssim90$ for non-relativistic electrons. Hence, for the gas
ionization fractions of $\lesssim 10^{-4}$, typical for molecular clouds, the contribution of Coulomb collisions should be
completely negligible.

\section{Spectrum of secondary electrons}
\label{spectrum_ch}

In this section we analyze generic properties of the secondary electron spectra $j_{\rm sec}(\varepsilon, N)$, related via
Equation~(\ref{F_aux}) to the solution of balance equation~(\ref{balance}). We consider only excitation of electronic
states; rotational and vibrational excitation, occurring at $\varepsilon<I$, are neglected. First, we derive the analytical
asymptotes valid for sufficiently high electron energies, and then compare this with exact numerical solution, which allows
us to elucidate the role of different inelastic processes in shaping the electron spectrum.

\subsection{Analytical solution at high energies}
\label{asym}

To evaluate the high-energy solution of Equation~(\ref{balance}), describing the electron spectrum at $\varepsilon\gg I$,
let us first neglect excitation collisions. As shown in Section~\ref{num}, their addition does not qualitatively affect the
results at high energies.

The right panel of Figure~\ref{fig_zeta12} demonstrates that the rates $\zeta_1$ and $\zeta_2$ are practically independent
of $\varepsilon$ for $\varepsilon\ll\varepsilon_{\rm att}(N)$, so that their sum is $\approx\zeta_{p}(N)$. On the other
hand, for the interstellar spectrum $\mathscr{H}$, both terms start rapidly decreasing at
$\varepsilon\gtrsim\varepsilon_{\rm att}$. Thus, for the analytical solution in this case it is reasonable to approximate
both $\zeta_1$ and $\zeta_2$ by step-functions, set to the respective ($N$-dependent) constants at
$\varepsilon\lesssim\varepsilon_{\rm att}$ and to zero at larger $\varepsilon$. We note that the step-function approximation
becomes exact for a monoenergetic local spectrum of protons with $E=E_{\rm att}(N)$.

For $\tilde\varepsilon$ much larger than $\eta_p\ln \tilde\varepsilon_{\rm att}$, the primary ionization in
Equation~(\ref{balance}) is dominated by the term $\propto\zeta_2$. In Appendix~\ref{A3} we derive the following leading
energy dependence for $\eta_p\ln \tilde \varepsilon_{\rm att} \ll \tilde \varepsilon \ll \tilde\varepsilon_{\rm att}$:
\begin{equation}\label{F_asy}
F(\varepsilon,N)\approx\frac{2\zeta_2(N)}{(1+\eta_e)I}\:
\frac{(\ln\tilde\varepsilon_{\rm att})^{\frac1{1+\eta_e}}}{\tilde\varepsilon\:(\ln\tilde\varepsilon)^{1+\frac1{1+\eta_e}}}\;,
\end{equation}
where the dependence on $N$ is given by Equation~(\ref{zeta_2}) evaluated at $\varepsilon=I$. In Appendix~\ref{A3} we also
obtain a rough estimate for the solution at lower energies, where the primary ionization is dominated by the term
$\propto\zeta_1$. Assuming $1\ll \tilde\varepsilon\ll \eta_p\ln \tilde\varepsilon_{\rm att}$ yields
\begin{equation}\label{F_asy2}
F(\varepsilon,N)\sim\frac{\zeta_1(N)}{(1+\eta_e)I}\:\frac1{\tilde\varepsilon^2}\;,
\end{equation}
with $\zeta_1(N)$ from Equation~(\ref{zeta_1}). This estimate neglects a factor $\sim1$, which logarithmically depends on
$\varepsilon$.

We remind that the physical spectrum of secondary electrons at $\tilde\varepsilon\gg 1$ scales as $j_{\rm
sec}(\varepsilon)\propto \varepsilon F(\varepsilon)$, as follows from Equation~(\ref{F_aux}). Therefore, the spectrum is
characterized by a long tail decreasing logarithmically with energy up to $\varepsilon\sim\varepsilon_{\rm att}(N)$. In
particular, this implies that the average energy of secondary electrons $\langle \varepsilon_{\rm sec} \rangle$ increases
with $N$; using Equation~(\ref{F_asy}), we readily obtain the following dependence:
\begin{equation}\label{eps_av}
\langle \varepsilon_{\rm sec} \rangle \approx\frac13\varepsilon_{\rm att}(N),
\end{equation}
derived assuming $\ln\tilde\varepsilon_{\rm att}\gg1$. It is important to stress that $\langle \varepsilon_{\rm sec}
\rangle$ is much larger than the average energy of electrons ejected in ionizing (primary or secondary) collisions, which is
generally calculated as $\langle \varepsilon_{\rm ej} \rangle= L(E)/\sigma_{\rm ion}(E)-I$. The latter is sometimes
erroneously employed in literature to characterize the average energy of secondary electrons. For primary ionization,
assuming proton energies $E\gg (m_p/m_e)I$, we can use Equation~(\ref{x-section_p}) for $\partial \sigma_p/\partial
\varepsilon$. Substituting this in Equation~(\ref{L_p2}), we derive the leading logarithmic term for the proton loss
function, $L_p(E)\approx f_p(E)I^2(1+\eta_p) \ln(\frac{m_e}{m_p}\tilde E)$. With the logarithmic term in the ionization
cross section $\sigma_{{\rm ion},p}(E)$ from Equation~(\ref{sigma_ion_p}), we obtain that the average energy of electrons
ejected by high-energy protons tends to $\langle \varepsilon_{{\rm ej},p} \rangle\to (1+2/\eta_p)I$; the same line of
arguments yields $\langle \varepsilon_{{\rm ej},e} \rangle\to (1+2/\eta_e)I$ for the secondary ionization. Thus, the average
energy of ejected electrons at large $N$ tends to a constant value of $\langle \varepsilon_{\rm ej} \rangle\sim 3I$ (since
$\eta_{p,e}\sim1$), while the average energy of the actual secondary spectrum follows Equation~(\ref{eps_av}).

Equations~(\ref{F_asy}) and (\ref{F_asy2}) can be extended to a case where excitation collisions are taken into account. For
large $\varepsilon$, the cross sections for the electron impact excitation of H$_2$ singlet states behave similar to the
ionization cross section \citep[see, e.g.,][]{Dalgarno1999,Janev2003}, i.e., their ratios tend to constant values. As shown
in Appendix~\ref{A3}, the solution in this case is still given by Equation~(\ref{F_asy}) with $\eta_e$ replaced by
\begin{equation}\label{eta}
\eta_e^*=\left[1+\frac12\sum_k\frac{\Delta_k}{I}\left(\frac{\sigma_{{\rm exc},k}}{\sigma_{{\rm ion},e}}\right)_\infty\right]\eta_e\;,
\end{equation}
where the cross section ratios are evaluated at $\varepsilon\to\infty$.

\subsection{Numerical solution and its analysis}
\label{num}

\begin{figure*}%[!h]
\begin{center}
\includegraphics[width=0.95\textwidth]{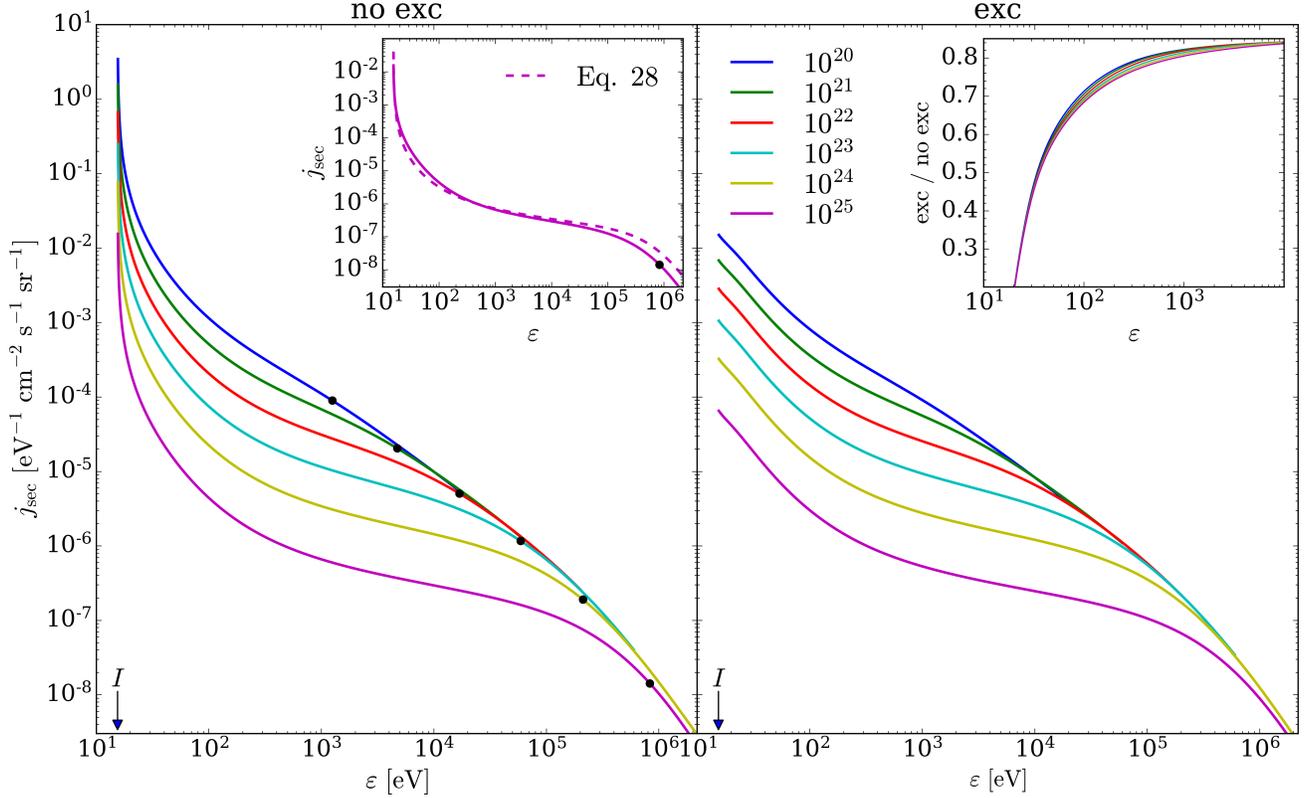}
\caption{Energy spectra of secondary electrons $j_{\rm sec}(\varepsilon)$ obtained for the interstellar proton spectrum
$\mathscr{H}$ from a numerical solution of Equation~(\ref{balance}) for different $N$ (see the legend, in units of
cm$^{-2}$). The vertical arrows in the bottom left corners of the panels indicate where $\varepsilon=I$. The left panel
shows the spectra for a model case where excitation collisions are omitted; agreement between the numerically computed
spectrum and the high-energy analytical asymptote, Equation~(\ref{F_asy}), is illustrated for $N=10^{25}$~cm$^{-2}$ in the
inset. The bullets for different curves indicate where $\varepsilon= \varepsilon_{\rm att}(N)$. The right panel displays the
results with excitation; the inset shows these spectra normalized by $j_{\rm sec}(\varepsilon)$ that are plotted in the left
panel.} \label{fig_spectrum}
\end{center}
\end{figure*}

The excitation cross sections of H$_2$ singlet and triplet states were taken from \citet{Janev2003}. The function $f_e(E)$,
relating $F$ and $j_{\rm sec}$ in Equation~(\ref{F_aux}), was derived from Equation~3 in \citet{Kim2000}.
Equation~(\ref{balance}) was solved numerically for discrete values of column between $N=10^{20}$~cm$^{-2}$ and
$N=10^{25}$~cm$^{-2}$, by implementing an iterative procedure for $F(\varepsilon)$ (similar to the solution of Volterra-type
integral equations). The next-iteration function $F_{i+1}(\varepsilon)$ was obtained by evaluating the lhs of
Equation~(\ref{balance}) for $F_i(\varepsilon)$, and then using this result to solve for $F_{i+1}(\varepsilon)$ on the rhs.
This procedure was repeated until $|F_{i+1}-F_i|\leq10^{-6} F_i$ at any $\varepsilon$. The convergence at smaller columns
was relatively fast and rather insensitive to the initial trial $F_0(\varepsilon)$. To facilitate the convergence at larger
columns, the initial trial for the next value of $N$ was the solution for the previous $N$.

Figure~\ref{fig_spectrum} displays $j_{\rm sec}(\varepsilon,N)$ computed for the interstellar proton spectrum $\mathscr{H}$.
The model case of no excitation is depicted in the left panel (``no exc'') by curves for different values of $N$. These
curves are well described by the high-energy analytical solution (\ref{F_asy}), as illustrated in the inset for
$N=10^{25}$~cm$^{-2}$. To facilitate the comparison with analytical results, we replaced the assumed step-function energy
dependence of $\zeta_2$ in Equation~(\ref{F_asy}) by the actual form determined by Equation~(\ref{zeta_2}) for the spectrum
$\mathscr{H}$. We see that the analytical curve in the inset remains accurate to within $30\%$ for $\varepsilon\gtrsim
I\ln\tilde\varepsilon_{\rm att}\:(\sim10^2$~eV) and $\varepsilon\ll\varepsilon_{\rm att}\:(\sim10^6$~eV). Remarkably, the
agreement remains reasonable (within a factor of 2--3) also for energies outside the assumed range of applicability.

Inclusion of excitation does not qualitatively change the form of $j_{\rm sec}(\varepsilon)$ except for energies in the
vicinity of the ionization potential, as evident from the right panel (``exc'') of Figure~\ref{fig_spectrum}. The inset
shows that excitation reduces $j_{\rm sec}(\varepsilon)$ by $\lesssim20\%$ at $\varepsilon\gtrsim1$~keV, almost irrespective
of $N$; the reduction is stronger at smaller $\varepsilon$, and the dependence on $N$ becomes more pronounced. This trend is
described by excitation correction~(\ref{eta}), leading to a reduction of high-energy spectra (\ref{F_asy}). Excitations of
H$_2$ singlet and triplet states contribute differently to this effect: cross sections $\sigma_{{\rm exc},k}(\varepsilon)$
for triplet states have a peak localized between 10--20~eV and rapidly decrease at larger $\varepsilon$, whereas for singlet
states they have a broader peak between $\sim$30--100~eV, overlapping with the peak of $\sigma_{{\rm ion},e}(\varepsilon)$,
and behave similarly to $\sigma_{{\rm ion},e}(\varepsilon)$ also at large $\varepsilon$. As discussed in the next section,
singlet excitations almost completely determine the magnitude of the secondary ionization rate, while the role of triplet
excitations is minor.

We note that the strong deviation seen between the left and right panels near the ionization potential originates from a
simple fact that, without excitation, the product $\sigma_{{\rm ion},e}(\varepsilon)j_{\rm sec}(\varepsilon)$ on the rhs of
Equation~(\ref{balance}) remains finite as $\varepsilon\to I$, thus leading to artificial divergence $j_{\rm
sec}(\varepsilon) \propto(\varepsilon-I)^{-1}$ in this case. This divergence does not significantly affect the calculation
of $\zeta_{\rm sec}$, because electrons with $\varepsilon\approx I$ provide a minor contribution to its value.

\section{Secondary ionization rate} \label{sec}

The rate of local secondary ionization can be conveniently rewritten in terms of the auxiliary functions $\Phi$ and $F$,
\begin{equation}\label{zeta_s}
\zeta_{\rm sec}(N)=\int_I^\infty\Phi(\varepsilon)F(\varepsilon,N)\:d\varepsilon.
\end{equation}
To obtain the ratio $\zeta_{\rm sec}/\zeta_{p}$ versus $N$, we derive $\zeta_{\rm sec}(N)$ by substituting the numerical
solution of Equation~(\ref{balance}), and calculate $\zeta_{p}(N)$ as explained in Section~\ref{primary}. For methodological
reasons, here we also discuss the model case of no excitation collisions -- this helps us to explore their impact on
$\zeta_{\rm sec}$ and to reveal the role of the interstellar proton spectrum. To characterize the effect of qualitatively
different proton spectra, here we present the results for both spectra $\mathscr{H}$ and $\mathscr{L}$.

\begin{figure}%[!h]
\begin{center}
\includegraphics[width=\columnwidth]{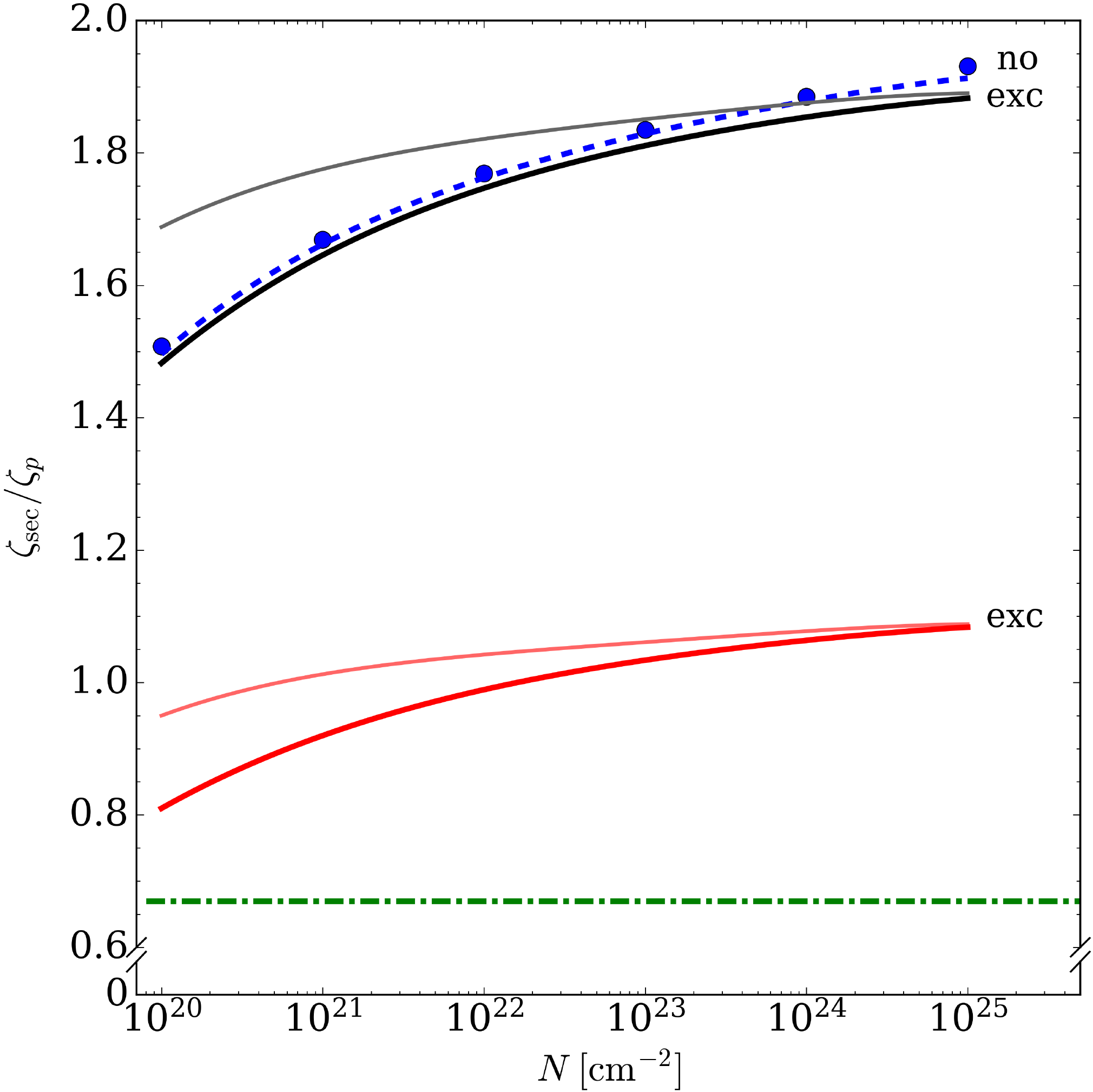}
\caption{Ratio of the secondary to primary ionization rates of H$_2$, $\zeta_{\rm sec}/\zeta_{p}$, as a function of
the gas column density, $N$, computed using the continuous slowing-down approximation for protons
(Section~\ref{CR_spectrum}) and a numerical solution of Equation~(\ref{balance}) for secondary electrons. Pairs of the
thicker and thinner solid lines represent results for the interstellar proton spectrum $\mathscr{H}$ and $\mathscr{L}$,
respectively. The pair of red curves (``exc'') shows the case where excitation collisions with H$_2$ are included. For
methodological reasons, we also plot the model case of no excitation (``no exc''). Here, in addition to the pair of black
curves the approximation of monoenergetic local protons with $E=E_{\rm att}(N)$ is also depicted: the dashed line represents
the numerical solution of Equation~(\ref{balance}), and the bullets show the results of Monte Carlo simulations. For
comparison, the horizontal dash-dotted line indicates the value of $\zeta_{\rm sec}/\zeta_{p} \approx0.67$ commonly adopted
in literature.} \label{fig_zeta_to_zeta}
\end{center}
\end{figure}

Figure~\ref{fig_zeta_to_zeta} summarizes our findings for $\zeta_{\rm sec}/\zeta_{p}$. We see that this ratio steadily
increases with column density: the trend is almost unaltered between the curves computed with and without excitation, and is
present for both proton spectra (though it is substantially weaker for the spectrum $\mathscr{L}$, see discussion below). In
Appendix~\ref{A4} it is shown that $\zeta_{\rm sec}/\zeta_{p}$ keeps increasing at any physically relevant value of $N$.
This behavior is quite different form the traditional assumption of a constant $\zeta_{\rm sec}/ \zeta_{p}$ with the
``canonical'' value of $\approx0.67$ \citep[e.g.,][]{Spitzer1968}.

We begin with the analysis of the results for the spectrum $\mathscr{H}$, shown by the {\it thick solid lines} in
Figure~\ref{fig_zeta_to_zeta}. The red line depicts $\zeta_{\rm sec}/\zeta_{p}$ versus $N$ for the ``exc'' case, where
excitation collisions are included. We see that this curve is shifted substantially down with respect to the model ``no
exc'' case (depicted by the black line), and that the slope of the ``exc'' curve is slightly smaller at larger $N$. This
behavior follows from the inset in the right panel of Figure~\ref{fig_spectrum}: excitation causes a reduction of $j_{\rm
sec}(\varepsilon)$ by 30--60\% at energies between $\sim$30--100~eV, corresponding to the maximum of the ionization cross
section (hence leading to an efficient decrease of $\zeta_{\rm sec}$), and the reduction is slightly stronger for larger
$N$. As noted in Section~\ref{num}, H$_2$ excitation is completely dominated by singlet states for
$\varepsilon\gtrsim30$~eV, and therefore the effect of triplet states on $\zeta_{\rm sec}/\zeta_{p}$ is minor: the ``exc''
curve computed for singlet excitation only would be shifted up by less than 10\% with respect to the curve shown in
Figure~\ref{fig_zeta_to_zeta}. Also, the results are virtually independent of the (poorly constrained) value of the
prefactor $\eta_e$ in Equation~(\ref{x-section_e}), varying by less than 1\% for $0.6\leq\eta_e\leq1$.

Let us now discuss on the role of the proton spectrum (considering for simplicity ``no exc'' case). For sufficiently soft
interstellar spectra, such as $\mathscr{H}$, the resulting local spectrum $j_{p}(E,N)$ is peaked at $E\sim E_{\rm att}(N)$,
as follows from Equation~(\ref{j_CR}). In Section~\ref{asym} we pointed out that this fact allows us to approximate the
rates of primary ionization $\zeta_{1,2}(\varepsilon,N)$ by step-functions of $\varepsilon$, which is equivalent to the
approximation of monoenergetic local protons with $E=E_{\rm att}(N)$. The dashed line in Figure~\ref{fig_zeta_to_zeta} shows
$\zeta_{\rm sec}/\zeta_{p}$ versus $N$ computed for this approximation, demonstrating a remarkably good agreement with the
corresponding thick solid line. To ensure an accurate comparison, $E_{\rm att}(N)$ was derived from the exact stopping range
of protons, as presented in \citet{Padovani2018}.

Thus, the dependence of $\zeta_{\rm sec}/\zeta_{p}$ on $N$ computed for the model spectrum $\mathscr{H}$ must be
representative of any sufficiently soft spectrum of interstellar protons. On the other hand, for extremely hard model
spectra -- such as $\mathscr{L}$, increasing at non-relativistic energies -- a monoenergetic approximation of local protons
is no longer justified. In this case, unattenuated protons with $E_{\rm att}(N)\lesssim E \lesssim E_0$ provide significant
contribution to primary ionization. The {\it thin solid lines} in Figure~\ref{fig_zeta_to_zeta} show $\zeta_{\rm
sec}/\zeta_{p}$ calculated for the spectrum $\mathscr{L}$, demonstrating that the resulting dependence on $N$ is noticeably
weaker than that for $\mathscr{H}$.

Available observational data on the H$_2$ ionization in a large number of diffuse clouds \citep{Indriolo2012,Neufeld2017}
tend to favor soft interstellar spectra. Assuming the continuous slowing-down approximation for CR protons, the spectrum
$\mathscr{H}$ provides a reasonable approximation of the data, while the spectrum $\mathscr{L}$ (which represents the
Voyager measurements, probing the very local ISM within the Local Bubble) underestimates the ionization rate in diffuse
clouds by more than an order of magnitude \citep{Padovani2018}. Moreover, the spectrum $\mathscr{L}$ fails to recover the
suggested dependence on $N$. Therefore, based on our current knowledge, one should consider the above results obtained for
the spectrum $\mathscr{H}$ as representative.

The fact that the approximation of monoenergetic local protons accurately describes the secondary ionization for soft
interstellar spectra allows us to substantiate and complement the above calculations by Monte Carlo simulations. In
Appendix~\ref{A5} we describe a simple algorithm to compute $\zeta_{\rm sec}/\zeta_{p}$ for monoenergetic protons directly,
based on the differential ionization cross sections given by Equations~(\ref{x-section_p}) and (\ref{x-section_e}). Results
of the direct simulations are depicted in Figure~\ref{fig_zeta_to_zeta} by the bullets, showing excellent agreement with the
dashed line.

Finally, adding interstellar CR nuclei heavier than protons does not significantly change the calculated values of
$\zeta_{\rm sec}/\zeta_{p}$. To estimate this effect, we keep in mind that the differential cross section of impact
ionization by a nucleus with the atomic number $Z$ is proportional to $Z^2$ and is determined by the nucleus velocity
\citep{Landau1981_quantum}. Assuming that Equation~(\ref{x-section_p}) describes the functional form of the differential
cross section for any nucleus, from Equation~(\ref{range0}) it follows that the attenuation energy per nucleon is equal to
$E_{\rm att}(Z^2N/A)$, where $A$ is the nucleus mass number. Hence, substituting $N\to Z^2N/A$ in a function describing the
dependence $\zeta_{\rm sec}/\zeta_{p}$ versus $N$ for protons, we obtain the corresponding dependence for nuclei. We employ
this fact in Appendix~\ref{A6} to show that the expected effect of heavier CR nuclei is to increase the ratios plotted in
Figure~\ref{fig_zeta_to_zeta} by less than 1\%.

\section{Conclusion and implications} \label{implications}

Our aim was to rigorously calculate the energy spectrum of secondary electrons that are produced by interstellar CRs
penetrating into dense regions of the ISM. The results are completely determined by the differential cross sections of the
proton impact (primary) and electron impact (secondary) ionization as well as by the electron excitation cross sections of
the gas species. We derived the governing balance equation which yields the secondary electron spectrum as a function of the
gas column density for a given regime of the proton penetration into dense gas; in this paper, the commonly used
free-streaming regime was assumed.

The principal findings can be summarized as follows:
\begin{enumerate}
\item The secondary electron spectrum $j_{\rm sec}(\varepsilon,N)$ has a long tail decreasing logarithmically with the
    energy $\varepsilon$, as described by the universal analytical asymptote~(\ref{F_asy}) for the auxiliary spectrum
    function $F(\varepsilon)\propto j_{\rm sec}(\varepsilon)/\varepsilon$. The effect of excitation collisions at high
    energies is generally described by Equation~(\ref{eta}).
\item The characteristic maximum energy of the secondary spectrum, $\varepsilon_{\rm att}(N)$, increases with the gas
    column $N$ according to Equation~(\ref{epsilon_att}). The maximum energy is proportional to the proton attenuation
    energy $E_{\rm att}(N)$, and the average energy of secondary electrons is $\sim\frac13\varepsilon_{\rm att}(N)$.
\item The ratio of the secondary to primary ionization rates, $\zeta_{\rm sec}/\zeta_{p}$, is a monotonically increasing
    function of the gas column for any relevant value of $N$. The value of $\zeta_{\rm sec}/\zeta_{p}$ varies between
    $\approx0.8$ and $\approx1.1$ for $10^{20}$~cm$^{-2}\leq N\leq 10^{25}$~cm$^{-2}$, as depicted by the thick red line
    in Figure~\ref{fig_zeta_to_zeta}, and thus is substantially larger than the commonly adopted constant value of
    $\approx0.67$.
\item The derived dependence $\zeta_{\rm sec}/\zeta_{p}$ versus $N$ is practically independent of a particular shape of
    the interstellar spectrum of protons (unless they have an extremely hard spectrum, such as the spectrum
    $\mathscr{L}$). This dependence can be accurately reproduced by using a monoenergetic local spectrum of protons with
    $E=E_{\rm att}(N)$.
\end{enumerate}

Knowing the actual form of the secondary electron spectrum opens up the possibility to accurately reevaluate characteristics
of several important processes driven by CRs in dark molecular clouds. The most notable and obvious examples include the gas
heating, production of atomic hydrogen, and generation of UV photons. It is certainly beyond the scope of this paper to
thoroughly analyze such processes, but we expect their characteristics to be significantly affected if the presented results
are taken into account, as outlined below:

{\it Gas heating.} Secondary electrons should contribute to the gas heating through additional ionization and excitation
channels. We can assess a relative energy budget for this process by comparing the rate of energy deposition due to
secondary ionization/excitation with that due to primary ionization/excitation \citep[but keeping in mind that only a
fraction of the energy deposited by CRs is eventually converted into heat, see][for detailed analysis]{Glassgold2012}.
Defining the deposition rate $\dot{\mathcal E}$ as the loss function averaged over the secondary and primary spectrum, the
ratio $\dot{\mathcal E}_{\rm sec}/\dot{\mathcal E}_p$ can be evaluated for large $N$ by virtue of Equation~(\ref{F_asy}),
similar to how we did it in Appendix~\ref{A4} for $\zeta_{\rm sec}/\zeta_{p}$. This yields the asymptotic ratio
$\dot{\mathcal E}_{\rm sec}/\dot{\mathcal E}_p\to \frac12(1+1/\eta_e)$, suggesting that the actual ratio of the heating
rates (i) may not be equal to $\zeta_{\rm sec}/\zeta_{p}$, as universally assumed in modeling, and (ii) may be sensitive to
the poorly constrained value of $\eta_e$.

{\it Production of atomic hydrogen.} Interstellar UV photons cannot penetrate the interiors of molecular clouds due to
absorption by dust as well as H$_2$ line absorption, and therefore the destruction of molecular hydrogen in these regions is
controlled by CRs. This process primarily occurs through electron-impact excitation of H$_2$ triplet states
\citep{Padovani2018b}, whose cross sections peak between 10--20~eV and rapidly decrease at higher energies. For this reason,
the rate of H$_2$ dissociation $\zeta_{\rm diss}$ must be particularly sensitive to the shape of the secondary electron
spectrum near the ionization potential. The secondary spectrum used to compute H$_2$ dissociation in \citet{Padovani2018b}
was derived from the continuous slowing-down approximation for electrons, leading to a practically constant ratio of
$\zeta_{\rm diss}/(\zeta_p+\zeta_{\rm sec})$ at columns of $N\gtrsim10^{21}$~cm$^{-2}$. Based on the results derived here
for $\zeta_{\rm sec}/\zeta_p$, we expect the ratio for H$_2$ dissociation to vary with $N$, too.

{\it Generation of UV photons.} Excitation of H$_2$ singlet states by CRs produces fluorescence in the Lyman and Werner
bands, leading to an efficient generation of UV field in dark clouds \citep{Prasad1983}. The cross section of electron
impact excitation of singlet states behaves similarly to the ionization cross section at energies above $\sim30$~eV, and
therefore the shape of the entire spectrum of secondary electrons is important for this process. Available estimates of the
UV field \citep{Cecchi-Pestellini1992} are also based on the continuous slowing-down approximation for electrons, assuming
the ``canonical'' value of $\zeta_{\rm sec}/\zeta_{p}\approx0.67$, and therefore one may expect significant corrections for
the UV field, too.

All three processes discussed above play an essential role in the physical and chemical evolution of molecular clouds, with
profound implications for the formation of stars and circumstellar disks.
%We plan to present their detailed analysis in separate publications.

We would like to thank Paola Caselli and Valerio Lattanzi for useful discussions and suggestions. A.V.I. acknowledges
support by the Russian Science Foundation via project 20-12-00047.

\appendix

\section{Appendix A: Differential form of energy losses for protons}
\label{A1}

Using Equation~(\ref{eps_max_p}), we obtain the upper integration limit $\varepsilon_{{\rm max},p}^*(E)$ in
Equation~(\ref{S_pr}),
\begin{equation}\label{eps_max_p*}
\varepsilon_{{\rm max},p}^*=\frac{\chi}{1-\chi}E-I,
\end{equation}
where $\chi\equiv4(m_e/m_p)$ is a small parameter characterizing the fraction of energy transferred to electrons. This
allows us to Taylor expand the integrand of Equation~(\ref{S_pr}) over small $\varepsilon+I$. Keeping the first two terms
yields
\begin{eqnarray}
\mathcal{P}(E)\approx j_{p}(E)\int_0^{\varepsilon_{{\rm max},p}^*}\frac{\partial \sigma_p}{\partial \varepsilon}(E,\varepsilon)\:
d\varepsilon \hspace{2.5cm}\label{S_pr2}\\
+\int_0^{\varepsilon_{{\rm max},p}^*}\frac{\partial}{\partial E} \left[\frac{\partial \sigma_p}{\partial
\varepsilon}(E,\varepsilon)j_{p}(E)\right](\varepsilon+I)\:d\varepsilon.\nonumber
\end{eqnarray}
From Equation~(\ref{eps_max_p*}) we derive
\begin{equation*}
\varepsilon_{{\rm max},p}^*-\varepsilon_{{\rm max},p}=\frac{\chi}{1-\chi} (\varepsilon_{{\rm max},p}+I),
\end{equation*}
where $\varepsilon_{{\rm max},p}$ is given by Equation~(\ref{eps_max_p}). Hence, the difference $\mathcal{P}- \mathcal{D}$
in Equation~(\ref{protons}) can be written with accuracy $O(\chi)$ as a sum of
\begin{equation}\label{term}
\chi(\varepsilon_{{\rm max},p}+I)\frac{\partial \sigma_p}{\partial \varepsilon}(E,\varepsilon_{{\rm max},p})j_{p}(E),
\end{equation}
and the second term in Equation~(\ref{S_pr2}). Since $\partial\varepsilon_{{\rm max},p}^*/\partial E\approx\chi$, this
second term can be written as a derivative over $E$ of the integral minus $\chi$ times the integrand taken at
$\varepsilon=\varepsilon_{{\rm max},p}^*$. To the same accuracy, the latter cancels out with term~(\ref{term}), and we
obtain
\begin{equation}\label{difference}
\mathcal{P}- \mathcal{D}=\frac{\partial}{\partial E}\left(L_pj_{p}\right)+O(\chi),
\end{equation}
where
\begin{equation}\label{L_p2}
L_p(E)=\int_0^{\varepsilon_{{\rm max},p}(E)}(\varepsilon+I)\frac{\partial \sigma_p}{\partial \varepsilon}(E,\varepsilon)\:
d\varepsilon,
\end{equation}
is the ionization loss function of protons.

\section{Appendix B: Analytical solution of Equation~(12)}
\label{A2}

An explicit solution of Equation~(\ref{protons2}) can be derived for a power-law interstellar spectrum,
\begin{equation}\label{j_CR0}
j_{p}^{\rm IS}(E)=j_0\left(\frac{E}{E_0}\right)^{-a}.
\end{equation}
Assuming CRs enter a cloud from one side ($\mu>0$), and substituting a power-law approximation of the proton loss function
\citep{Padovani2018,Silsbee2019},
\begin{equation}\label{L_p}
L_p(E)=L_0\left(\frac{E}{E_0}\right)^{-d},
\end{equation}
valid for $4\times10^5$~eV~$\lesssim E\lesssim 10^8$~eV (with $d=0.81$ and $L_0 =1.21\times10^{-17}$~eV~cm$^2$ for
$E_0=650$~MeV), we obtain the following solution \citep[see Appendix~E of][]{Padovani2018}:
\begin{equation}\label{j_CR}
j_{p}(E,N,\mu)=j_{p}^{\rm IS}(E)\left[1+\frac{N}{\mu R_p(E)}\right]^{-\frac{a+d}{1+d}},
\end{equation}
where
\begin{equation}\label{R_p}
R_p(E)=\frac{E_0}{(1+d)L_0}\left(\frac{E}{E_0}\right)^{1+d},
\end{equation}
is the proton stopping range for the loss function~(\ref{L_p}).

\section{Appendix C: High-energy spectrum of secondary electrons}
\label{A3}

Let us first omit excitation collisions. In order to evaluate the high-energy spectrum at $\varepsilon\gg I$, we break the
integrals on the lhs of Equation~(\ref{balance}) into two parts: from 0 to $\varepsilon$ (``integrals~I''), and from
$\varepsilon$ to $\infty$ (``integrals~II''). Since $\zeta_{1,2}$ rapidly decrease at $\varepsilon\gtrsim\varepsilon_{\rm
att}$, we assume that $F(\varepsilon)$ vanishes at these energies. Below it is shown that the leading term of the
high-energy spectrum depends logarithmically on $\varepsilon_{\rm att}$, and therefore we can truncate integrals~II at
$\varepsilon'=\varepsilon_{\rm att}$.

We substitute $\varphi_{\rm M}(\varepsilon,\varepsilon')$ and $\varphi_{\rm dip}(\varepsilon,\varepsilon')$ from
Equations~(\ref{Mott}) and (\ref{dip}) into the integrals and neglect unity in the terms containing $1+\tilde\varepsilon$.
Then we {\it multiply} Equation~(\ref{balance}) by $\tilde\varepsilon^2$ and write the resulting sum of integrals~II in the
following form:
\begin{eqnarray}
\sum\int_{\varepsilon}^{\varepsilon_{\rm att}}\ldots\approx\int_{2\varepsilon}^{\varepsilon_{\rm att}}F(\varepsilon')\Bigg\{1-
\frac{\varepsilon}{\varepsilon'-\varepsilon}\hspace{2.2cm}\label{Sum2}\\[.1cm]
+\left(\frac{\varepsilon}{\varepsilon'-\varepsilon}\right)^2+\eta_e\frac{\ln\tilde\varepsilon'}{\tilde\varepsilon}
\left[1+\left(\frac{\varepsilon}{\varepsilon'-\varepsilon}\right)^3\right]\Bigg\}\:d\varepsilon'.\nonumber
\end{eqnarray}
We see that the small terms $\propto \ln\tilde\varepsilon'/\tilde\varepsilon$ can be safely neglected. For the sum of
integrals~I, we obtain
\begin{widetext}
\begin{eqnarray}
\sum\int_0^{\varepsilon}\ldots\approx\int_0^{\varepsilon}F(\varepsilon+\varepsilon'+I)\:d\varepsilon'-
\tilde\varepsilon\int_0^{\varepsilon}\frac{F(\varepsilon+\varepsilon'+I)}{1+\tilde\varepsilon'}\:d\varepsilon'+
\tilde\varepsilon^2\int_0^{\varepsilon}\frac{F(\varepsilon+\varepsilon'+I)}{(1+\tilde\varepsilon')^2}\:d\varepsilon'
\hspace{3cm}\label{Sum1}\\[.3cm]
+\frac{\eta_e}{\tilde\varepsilon}\int_0^{\varepsilon}\ln(\tilde\varepsilon+\tilde\varepsilon'+1)
F(\varepsilon+\varepsilon'+I)\:d\varepsilon' +\eta_e\tilde\varepsilon^2\int_0^{\varepsilon}
\frac{\ln(\tilde\varepsilon+\tilde\varepsilon'+1)}{(1+\tilde\varepsilon')^3}\:F(\varepsilon+\varepsilon'+I)\:d\varepsilon'.
\nonumber
\end{eqnarray}
\end{widetext}
We Taylor expand $F(\varepsilon+\varepsilon'+I)$ and $\ln(\tilde\varepsilon+\tilde\varepsilon'+1)$ over
$\tilde\varepsilon'+1$. Terms $\propto F(\varepsilon)$ include those leading in $\varepsilon$ (from the second, third, and
fifth integrals~I) which exactly cancel out with the rhs terms. Keeping the remaining leading terms resulting from the
expansion yields
\begin{eqnarray}
\frac1I\sum\int_0^{\varepsilon}\ldots\approx (1+\eta_e)\tilde\varepsilon\:F(\tilde\varepsilon)\hspace{3.5cm}\label{Sum1a}\\
\textstyle +(1+\eta_e) \tilde\varepsilon^2 \ln\tilde\varepsilon\: F'(\tilde\varepsilon)+\frac5{12}\tilde\varepsilon^3\:
F''(\tilde\varepsilon)+\ldots,\nonumber
\end{eqnarray}
where $F'(\tilde\varepsilon)$ denotes the derivative with respect to $\tilde\varepsilon$. {\it A posteriori} analysis
renders terms with the second and higher derivatives in Equation~(\ref{Sum1a}) unimportant for the leading term of the
sought solution.

For $\tilde\varepsilon\gg \eta_p\ln \tilde\varepsilon_{\rm att}$, the leading term due to primary ionization is $2\zeta_2$
[in the multiplied Equation~(\ref{balance})]. Summing up, for $\eta_p\ln \tilde\varepsilon_{\rm att}\ll \tilde\varepsilon
\ll \tilde\varepsilon_{\rm att}$ Equation~(\ref{balance}) is reduced to
\begin{eqnarray}
(1+\eta_e)\left[\tilde\varepsilon^2\ln\tilde\varepsilon\: \tilde F'(\tilde\varepsilon)+
\tilde\varepsilon\: \tilde F(\tilde\varepsilon)\right]\hspace{3.4cm}\label{balance2}\\[.1cm]
+\int_{2\tilde\varepsilon}^{\tilde\varepsilon_{\rm att}}\tilde F(\tilde\varepsilon')
\left[1-\frac1{\tilde\varepsilon'/\tilde\varepsilon-1}+\frac1{(\tilde\varepsilon'/\tilde\varepsilon-1)^2}\right]d\tilde
\varepsilon'+1=0,\nonumber
\end{eqnarray}
where $\tilde F=(I/2\zeta_2)F$. We introduce a new variable $x=\ln\tilde\varepsilon$ and seek the solution of the form
$\tilde F(\tilde \varepsilon)=ce^{-x}/ x^{1+s}$. As the leading contribution of the first term, $\propto 1/x^s$, is provided
by the first term in the brackets, we obtain the following equation:
\begin{eqnarray}
-(1+\eta_e)\frac{c}{x^s}\hspace{6.6cm}\label{balance3}\\[.2cm]
+c\int_{x+\ln2}^{x_{\rm att}}\left[1-\frac1{e^{x'-x}-1}+\frac1{(e^{x'-x}-1)^2}\right]\frac{dx'}{x'^{1+s}}+1=0,\nonumber
\end{eqnarray}
with $x_{\rm att}=\ln\tilde \varepsilon_{\rm att}$. The integral term yields $(c/s)(1/x^s-1/x_{\rm att}^s)+O(1/x^{1+s})$. We
see that Equation~(\ref{balance3}) is satisfied for $s=(1+\eta_e)^{-1}$ and $c=sx_{\rm att}^s$, which gives us
Equation~(\ref{F_asy}).

One can also roughly estimate the form of electron spectrum at $\tilde\varepsilon\lesssim \eta_p\ln \tilde\varepsilon_{\rm
att}$, still assuming $\tilde\varepsilon\gg1$. In this case, the primary ionization in Equation~(\ref{balance}) is dominated
by the term $2\zeta_1/\tilde\varepsilon^3$. Keeping in mind that the above analysis is performed for the term
$2\zeta_2/\tilde\varepsilon^2$, we conclude that the sought spectrum obeys Equation~(\ref{balance2}) with the last term
(unity) replaced by $1/\tilde\varepsilon$. One can see that, up to a factor depending on $\ln\tilde\varepsilon$, the
solution is given by the following leading term:
\begin{equation}\label{balance4}
\tilde F(\varepsilon)\sim\frac1{2(1+\eta_e)\tilde\varepsilon^2}\;,
\end{equation}
where $\tilde F=(I/2\zeta_1)F$. This gives us Equation~(\ref{F_asy2}).

To include the contribution of excitation collisions to the high-energy solution, we take into account that cross sections
for H$_2$ ionization and excitation (of singlet states) behave similarly at large $\varepsilon$
\citep[][]{Dalgarno1999,Janev2003}. This implies that the ratio $\sigma_{{\rm exc},k}/\sigma_{{\rm ion},e}$ tends to a
constant as $\varepsilon\to\infty$. Then, expanding a difference of the excitation terms in Equation~(\ref{balance}) and
taking into account Equation~(\ref{sigma_ion_e}) yields the following additional contribution to Equation~(\ref{balance2}):
\begin{equation*}
\frac{\eta_e}2\left[\tilde\varepsilon^2\ln\tilde\varepsilon\: \tilde F'(\tilde\varepsilon)+
\tilde\varepsilon\: \tilde F(\tilde\varepsilon)\right]
\sum_k\frac{\Delta_k}{I}\left(\frac{\sigma_{{\rm exc},k}}{\sigma_{{\rm ion},e}}\right)_{\infty},
\end{equation*}
where the cross section ratios are evaluated at $\varepsilon\to\infty$. Hence, high-energy asymptote (\ref{F_asy}) is valid
also in the presence of excitation collisions, where $\eta_e$ should be replaced with modified value $\eta_e^*$ given by
Equation~(\ref{eta}).

\section{Appendix D: Analytical estimates of $\zeta_{\rm sec}/\zeta_{p}$}
\label{A4}

The high-energy analytical spectrum of secondary electrons, given by Equations~(\ref{F_asy}) and (\ref{F_asy2}), allows us
to qualitatively understand Figure~\ref{fig_zeta_to_zeta}.

For monoenergetic protons with $E=E_{\rm att}(N)$, the energy dependence of $\zeta_1$ and $\zeta_2$ can be approximated by
step functions, and $\zeta_{p}=[1+ \eta_p\ln(\tilde \varepsilon_{\rm att}/4)]\zeta_2$. Let us write $\zeta_{\rm sec}$ as a
sum $\zeta_{\rm sec1}+\zeta_{\rm sec2}$, representing contributions $F\propto \zeta_{1,2}$ to the integral in
Equation~(\ref{zeta_s}). We calculate $\zeta_{\rm sec2}$ by substituting Equation~(\ref{F_asy}) and integrating over
$\tilde\varepsilon$ from $\eta_p\ln \tilde\varepsilon_{\rm att}$ to $\tilde\varepsilon_{\rm att}$; as Equation~(\ref{F_asy})
gives the leading dependence on $\ln\tilde \varepsilon$, we keep only the leading term also in Equation~(\ref{sigma_ion_e})
and substitute $\Phi(\varepsilon)\approx \frac12\eta_e \ln\tilde\varepsilon$. This yields the asymptotic expression
\begin{equation}\label{zeta_r}
\frac{\zeta_{\rm sec2}}{\zeta_{p}}\approx\frac1{\eta_p}-\frac1{\eta_p}
\left(\frac{\ln\tilde\varepsilon_{\rm att}}{\ln\ln\tilde\varepsilon_{\rm att}}\right)^{-\frac{\eta_e}{1+\eta_e}},
\end{equation}
which is formally valid for sufficiently large $N$ and neglects further corrections depending on $\ln\ln\tilde
\varepsilon_{\rm att}$. As regards $\zeta_{\rm sec1}$, Equation~(\ref{F_asy2}) is too crude to obtain a quantitative
estimate. Nevertheless, it allows us to understand how $\zeta_{\rm sec1}$ depends on $\varepsilon_{\rm att}$. For large
$\tilde\varepsilon_{\rm att}$ one can write $\zeta_1\approx\zeta_{p}$. Substituting Equation~(\ref{F_asy2}) and
$\Phi(\varepsilon)$ determined by Equation~(\ref{sigma_ion_e}) in Equation~(\ref{zeta_s}) and integrating from $\sim1$ to
$\sim\ln \tilde \varepsilon_{\rm att}$ gives
\begin{equation}\label{zeta_r2}
\frac{\zeta_{\rm sec1}}{\zeta_{p}}\sim {\rm const}-\frac1{\ln\tilde\varepsilon_{\rm att}}\;,
\end{equation}
with const~$\approx0.4$. The logarithmic dependence of the integrand on $\varepsilon$, omitted to derive
Equation~(\ref{zeta_r2}), results in terms $\ln\ln\tilde \varepsilon_{\rm att}$. The convergence of Equation~(\ref{zeta_r2})
to a constant at $N\to\infty$ is substantially faster than that of Equation~(\ref{zeta_r}), and therefore the latter
determines the behavior of $\zeta_{\rm sec}/\zeta_{p}$ at large $N$.

We note that the asymptotic convergence of $\zeta_{\rm sec}/\zeta_{p}$ is very slow -- it actually occurs at unphysically
large column densities (of $\gtrsim10^{30}$~cm$^{-2}$). Hence, Equation~(\ref{zeta_r}) merely serves as an indicator that
$\zeta_{\rm sec}/\zeta_{p}$ keeps increasing at any relevant $N$.

\section{Appendix E: Monte Carlo simulations for monoenergetic protons}
\label{A5}

Using a step-function model for $\zeta_{1,2}$ -- which assumes that the primary ionization at given $N$ is produced by
protons with the energy $E_{\rm att}(N)$ -- enables an easy comparison with direct simulations of the secondary ionization.
The simulation algorithm is as follows: The first-generation secondary electron acquires the energy $\varepsilon_1$ in the
range from 0 to $4(m_e/m_p)E_{\rm att}(N)-I$, with the probability given by Equation~(\ref{x-section_p}). Then, if that
electron has energy greater than $I$, it creates the second-generation secondary electron with the energy $\varepsilon_2$ in
the range from 0 to $\frac12(\varepsilon_1-I)$, with the probability given by Equation~(\ref{x-section_e}); simultaneously,
the energy of the first-generation electron is reduced from $\varepsilon_1$ to $\varepsilon_1- \varepsilon_2-I$. This
process is repeated with all electrons with energy greater than $I$ until there are none remaining. Then $\zeta_{\rm
sec}/\zeta_{p}$ is equal to the average number of electrons in the simulation minus one, obtained after averaging over
$10^9$ primary ionizations.

\section{Appendix F: Effect of heavier CR nuclei}
\label{A6}

Let us denote by $\mathcal{R}(N)$ a dependence $\zeta_{\rm sec}/\zeta_{p}$ versus $N$ due to CR protons (one of the curves
depicted in Figure~\ref{fig_zeta_to_zeta}). As explained in Section~\ref{sec}, the corresponding dependence for heavier
nuclei $k$, with the atomic number $Z_k$ and mass number $A_k$, is given by $\zeta_{\rm
sec}^{(k)}/\zeta_{p}^{(k)}=\mathcal{R}(\psi_kN)$, where
\begin{equation}
\psi_k=Z_k^2/A_k\;,
\end{equation}
is unity for $^4$He and $>1$ for other stable nuclei. Our aim is to calculate the ratio of the total secondary ionization
rate, $\zeta_{\rm sec}^{\Sigma}= \zeta_{\rm sec}+\sum_k\zeta_{\rm sec}^{(k)}$, to the total primary rate,
$\zeta_{p}^{\Sigma}= \zeta_{p}+ \sum_k \zeta_{p}^{(k)}$. Simple manipulation yields
\begin{equation}\label{zeta_sum}
\frac{\zeta_{\rm sec}^{\Sigma}}{\zeta_{p}^{\Sigma}}=\mathcal{R}(N)\:\frac{1+\sum_k\frac{\mathcal{R}(\psi_kN)}{\mathcal{R}(N)}
\frac{\zeta_{p}^{(k)}}{\zeta_{p}}}{1+\sum_k\frac{\zeta_{p}^{(k)}}{\zeta_{p}}}\;.
\end{equation}
The ratio of the primary ionization rates, $\zeta_{p}^{(k)}/\zeta_{p}$, is evaluated by employing the dependence
$\zeta_p(N)\propto N^{-\alpha}$ with $\alpha=\frac{a+d-1}{1+d}\:$, derived for a power-law interstellar spectrum of protons
by \citet{Silsbee2019} (see also Appendix~\ref{A2}). Given that the attenuation energy (per nucleon) of nucleus $k$ is
$E_{\rm att}(\psi_kN)$ and that their ionization rate is proportional to $Z_k^2\equiv A_k\psi_k$, we obtain
\begin{equation}\label{zeta_p_ratio}
\frac{\zeta_{p}^{(k)}}{\zeta_{p}}=x_kA_k\psi_k^{1-\alpha},
\end{equation}
where $x_k$ is the interstellar abundance relative to protons.

Equation~(\ref{zeta_sum}) shows that the effect of heavier nuclei is to increase the relative magnitude of the secondary
ionization, because $\mathcal{R}(N)$ is an increasing function. However, the magnitude of the effect is negligible. Assuming
the interstellar spectrum $\mathscr{H}$ with $a=0.8$, which gives $\alpha\approx0.35$, and using galactic CR abundances
estimated from \citet{Dartois2015}, we conclude that the fraction factor on the rhs of Equation~(\ref{zeta_sum}) differs
form unity by less than 1\%.

\bibliographystyle{apj}
\bibliography{paper}

\begin{thebibliography}{}
\expandafter\ifx\csname natexlab\endcsname\relax\def\natexlab#1{#1}\fi

\bibitem[{{Bethe}(1930)}]{Bethe1930}
{Bethe}, H. 1930, Annalen der Physik, 397, 325

\bibitem[{{Caselli} \& {Ceccarelli}(2012)}]{Caselli2012}
{Caselli}, P., \& {Ceccarelli}, C. 2012, \aapr, 20, 56

\bibitem[{{Caselli} {et~al.}(1998){Caselli}, {Walmsley}, {Terzieva}, \&
  {Herbst}}]{Caselli1998}
{Caselli}, P., {Walmsley}, C.~M., {Terzieva}, R., \& {Herbst}, E. 1998, \apj,
  499, 234

\bibitem[{{Cecchi-Pestellini} \& {Aiello}(1992)}]{Cecchi-Pestellini1992}
{Cecchi-Pestellini}, C., \& {Aiello}, S. 1992, \mnras, 258, 125

\bibitem[{{Cravens} \& {Dalgarno}(1978)}]{Cravens1978}
{Cravens}, T.~E., \& {Dalgarno}, A. 1978, \apj, 219, 750

\bibitem[{{Cummings} {et~al.}(2016){Cummings}, {Stone}, {Heikkila}, {Lal},
  {Webber}, {J{\'o}hannesson}, {Moskalenko}, {Orlando}, \&
  {Porter}}]{Cummings2016}
{Cummings}, A.~C., {Stone}, E.~C., {Heikkila}, B.~C., {et~al.} 2016, \apj, 831,
  18

\bibitem[{{Dalgarno} \& {Griffing}(1958)}]{Dalgarno1958}
{Dalgarno}, A., \& {Griffing}, G.~W. 1958, Proceedings of the Royal Society of
  London Series A, 248, 415

\bibitem[{{Dalgarno} {et~al.}(1999){Dalgarno}, {Yan}, \& {Liu}}]{Dalgarno1999}
{Dalgarno}, A., {Yan}, M., \& {Liu}, W. 1999, \apjs, 125, 237

\bibitem[{{Dartois} {et~al.}(2015){Dartois}, {Aug{\'e}}, {Rothard}, {Boduch},
  {Brunetto}, {Chabot}, {Domaracka}, {Ding}, {Kamalou}, {Lv}, {da Silveira},
  {Thomas}, {Pino}, {Mejia}, {Godard}, \& {de Barros}}]{Dartois2015}
{Dartois}, E., {Aug{\'e}}, B., {Rothard}, H., {et~al.} 2015, Nuclear
  Instruments and Methods in Physics Research B, 365, 472

\bibitem[{{Erskine}(1954)}]{Erskine1954}
{Erskine}, G.~A. 1954, Proceedings of the Royal Society of London Series A,
  224, 362

\bibitem[{{Fano}(1953)}]{Fano1953}
{Fano}, U. 1953, Physical Review, 92, 328

\bibitem[{{Galli} {et~al.}(2002){Galli}, {Walmsley}, \&
  {Gon{\c{c}}alves}}]{Galli2002}
{Galli}, D., {Walmsley}, M., \& {Gon{\c{c}}alves}, J. 2002, \aap, 394, 275

\bibitem[{{Glassgold} {et~al.}(2012){Glassgold}, {Galli}, \&
  {Padovani}}]{Glassgold2012}
{Glassgold}, A.~E., {Galli}, D., \& {Padovani}, M. 2012, \apj, 756, 157

\bibitem[{{Glassgold} \& {Langer}(1973)}]{Glassgold1973}
{Glassgold}, A.~E., \& {Langer}, W.~D. 1973, \apj, 186, 859

\bibitem[{{Indriolo} \& {McCall}(2012)}]{Indriolo2012}
{Indriolo}, N., \& {McCall}, B.~J. 2012, \apj, 745, 91

\bibitem[{{Ivlev} {et~al.}(2019){Ivlev}, {Silsbee}, {Sipil{\"a}}, \&
  {Caselli}}]{Ivlev2019}
{Ivlev}, A.~V., {Silsbee}, K., {Sipil{\"a}}, O., \& {Caselli}, P. 2019, \apj,
  884, 176

\bibitem[{{Janev} {et~al.}(2003){Janev}, {Reiter}, \& {Samm}}]{Janev2003}
{Janev}, R.~K., {Reiter}, D., \& {Samm}, U. 2003, {Collision Processes in
  Low-Temperature Hydrogen Plasmas} (J{\"u}lich: Forschungszentrum,
  Zentralbibliothek)

\bibitem[{{Keto} \& {Caselli}(2008)}]{Keto2008}
{Keto}, E., \& {Caselli}, P. 2008, \apj, 683, 238

\bibitem[{{Keto} {et~al.}(2014){Keto}, {Rawlings}, \& {Caselli}}]{Keto2014}
{Keto}, E., {Rawlings}, J., \& {Caselli}, P. 2014, \mnras, 440, 2616

\bibitem[{{Kim} \& {Rudd}(1994)}]{Kim1994}
{Kim}, Y.-K., \& {Rudd}, M.~E. 1994, \pra, 50, 3954

\bibitem[{{Kim} {et~al.}(2000){Kim}, {Santos}, \& {Parente}}]{Kim2000}
{Kim}, Y.-K., {Santos}, J.~P., \& {Parente}, F. 2000, \pra, 62, 052710

\bibitem[{{Knipp} {et~al.}(1953){Knipp}, {Eguchi}, {Ohta}, \&
  {Nagata}}]{Knipp1953}
{Knipp}, J.~K., {Eguchi}, T., {Ohta}, M., \& {Nagata}, S. 1953, Progress of
  Theoretical Physics, 10, 24

\bibitem[{Landau \& Lifshitz(1991)}]{Landau1981_quantum}
Landau, L., \& Lifshitz, E. 1991, Quantum Mechanics: Non-Relativistic Theory
  (Oxford: Pergamon)

\bibitem[{{McKee}(1989)}]{McKee1989}
{McKee}, C.~F. 1989, \apj, 345, 782

\bibitem[{{Mott}(1930)}]{Mott1930}
{Mott}, N.~F. 1930, Proceedings of the Royal Society of London Series A, 126,
  259

\bibitem[{{Neufeld} \& {Wolfire}(2017)}]{Neufeld2017}
{Neufeld}, D.~A., \& {Wolfire}, M.~G. 2017, \apj, 845, 163

\bibitem[{{Padovani} {et~al.}(2018{\natexlab{a}}){Padovani}, {Galli}, {Ivlev},
  {Caselli}, \& {Ferrara}}]{Padovani2018b}
{Padovani}, M., {Galli}, D., {Ivlev}, A.~V., {Caselli}, P., \& {Ferrara}, A.
  2018{\natexlab{a}}, \aap, 619, A144

\bibitem[{{Padovani} {et~al.}(2018{\natexlab{b}}){Padovani}, {Ivlev}, {Galli},
  \& {Caselli}}]{Padovani2018}
{Padovani}, M., {Ivlev}, A.~V., {Galli}, D., \& {Caselli}, P.
  2018{\natexlab{b}}, \aap, 614, A111

\bibitem[{{Padovani} {et~al.}(2020){Padovani}, {Ivlev}, {Galli}, {Offner},
  {Indriolo}, {Rodgers-Lee}, {Marcowith}, {Girichidis}, {Bykov}, \&
  {Kruijssen}}]{Padovani2020}
{Padovani}, M., {Ivlev}, A.~V., {Galli}, D., {et~al.} 2020, \ssr, 216, 29

\bibitem[{{Prasad} \& {Tarafdar}(1983)}]{Prasad1983}
{Prasad}, S.~S., \& {Tarafdar}, S.~P. 1983, \apj, 267, 603

\bibitem[{{Rudd}(1987)}]{Rudd1987}
{Rudd}, M.~E. 1987, Radiation Research, 109, 1

\bibitem[{{Rudd}(1988)}]{Rudd1988}
---. 1988, \pra, 38, 6129

\bibitem[{{Rudd} {et~al.}(1992){Rudd}, {Kim}, {Madison}, \& {Gay}}]{Rudd1992}
{Rudd}, M.~E., {Kim}, Y.~K., {Madison}, D.~H., \& {Gay}, T.~J. 1992, Reviews of
  Modern Physics, 64, 441

\bibitem[{{Shu} {et~al.}(1987){Shu}, {Adams}, \& {Lizano}}]{Shu1987}
{Shu}, F.~H., {Adams}, F.~C., \& {Lizano}, S. 1987, \araa, 25, 23

\bibitem[{{Silsbee} \& {Ivlev}(2019)}]{Silsbee2019}
{Silsbee}, K., \& {Ivlev}, A.~V. 2019, \apj, 879, 14

\bibitem[{{Spencer} \& {Fano}(1954)}]{Spencer1954}
{Spencer}, L.~V., \& {Fano}, U. 1954, Physical Review, 93, 1172

\bibitem[{{Spitzer} \& {Tomasko}(1968)}]{Spitzer1968}
{Spitzer}, Lyman, J., \& {Tomasko}, M.~G. 1968, \apj, 152, 971

\bibitem[{{Wilms} {et~al.}(2000){Wilms}, {Allen}, \& {McCray}}]{Wilms2000}
{Wilms}, J., {Allen}, A., \& {McCray}, R. 2000, \apj, 542, 914

\bibitem[{{Xu} \& {McCray}(1991)}]{Xu1991}
{Xu}, Y., \& {McCray}, R. 1991, \apj, 375, 190

\bibitem[{{Zhao} {et~al.}(2018){Zhao}, {Caselli}, {Li}, \&
  {Krasnopolsky}}]{Zhao2018}
{Zhao}, B., {Caselli}, P., {Li}, Z.-Y., \& {Krasnopolsky}, R. 2018, \mnras,
  473, 4868

\bibitem[{{Zhao} {et~al.}(2016){Zhao}, {Caselli}, {Li}, {Krasnopolsky},
  {Shang}, \& {Nakamura}}]{Zhao2016}
{Zhao}, B., {Caselli}, P., {Li}, Z.-Y., {et~al.} 2016, \mnras, 460, 2050

\end{thebibliography}

\end{document}